\documentclass[10pt, journal]{IEEEtran}
\usepackage{graphicx}
\usepackage{caption}
\usepackage[ruled]{algorithm2e}

\usepackage{indentfirst}
\usepackage{graphicx}              
\usepackage{subcaption}            
\usepackage{graphicx}
\usepackage{bm}
\usepackage{amsmath}
\allowdisplaybreaks[4]
\usepackage{amssymb}
\usepackage{times}
\usepackage{mathtools}
\usepackage{psfrag}
\usepackage{cite}
\usepackage{lastpage}
\usepackage{fancyhdr}
\usepackage{color}
\usepackage{amsthm}

\usepackage{bigints}
\usepackage{multirow}
\usepackage{booktabs}
\usepackage{threeparttable}

\newtheorem{Theorem}{Theorem}

\newtheorem{Lemma}{Lemma}
\theoremstyle{remark}
\newtheorem{Remark}{$\mathbf{Remark}$}
\newtheorem{Corollary}{Corollary}
\newtheorem{theorem}[Theorem]{$\mathbf{Theorem}$}
\newtheorem{lemma}[Lemma]{$\mathbf{Lemma}$}

\newtheorem{corollary}[Corollary]{$\mathbf{Corollary}$}

\begin{document}
\title{An Energy Efficient Design of Hybrid NOMA Based on Hybrid SIC with Power Adaptation}
\author{
Ning Wang, Chenyu Zhang, Yanshi Sun,~\IEEEmembership{\itshape Member, IEEE}, Minghui Min,

Yuanwei Liu,~\IEEEmembership{{\itshape Fellow, IEEE}}, and Shiyin Li
\vspace{-1em}
\thanks{N. Wang, C. Zhang, M. Min and S. Li are with the School of Information and Control Engineering, China University of Mining and Technology, Xuzhou, 221116, China. (email: wangnsky@cumt.edu.cn, zcy6@cumt.edu.cn, minmh@cumt.edu.cn, lishiyin@cumt.edu.cn). 

Y. Sun is with the School of Computer Science and Information Engineering, Hefei University of Technology, Hefei, 230009, China. (email: sys@hfut.edu.cn). 

Y. Liu is with the Department of Electrical and Electronic Engineering, The University of Hong Kong, Hong Kong (email: yuanwei@hku.hk).}
}

\maketitle
\vspace{-2em}
\begin{abstract}
Recently, hybrid non-orthogonal multiple access (H-NOMA) technology, which effectively utilizes both NOMA and orthogonal multiple access (OMA) technologies through flexible resource allocation in a single transmission, has demonstrated immense potential for enhancing the performance of wireless communication systems. To further release the potential of H-NOMA,
this paper proposes a novel design of H-NOMA which jointly incorporates hybrid successive interference cancellation (HSIC) and power adaptation (PA) in the NOMA transmission phase. 
To reveal the potential of the proposed HSIC-PA aided H-NOMA scheme, closed-form expression
for the probability of the event that H-NOMA can achieve a higher data rate than pure OMA
by consuming less energy is rigorously derived. 
{\color{black} Furthermore, the asymptotic analysis demonstrates that the probability of the proposed H-NOMA scheme approaches $1$ in the high signal-to-noise ratio (SNR) regime without any constraints on either users' target rates or transmit power ratios. This represents a significant improvement over conventional H-NOMA schemes, which require specific restrictive conditions to achieve probability 1 at high SNRs as shown in existing work.} The above observation indicates that with less energy consumption, the proposed HSIC-PA aided H-NOMA can achieve a higher data rate than pure OMA with probability $1$ at high SNRs, and hence a higher energy efficiency. Finally, numerical results are provided to verify the accuracy of the analysis and also demonstrate
the superior performance of the proposed H-NOMA scheme.
\end{abstract}

\begin{IEEEkeywords}
Hybrid non-orthogonal multiple access (H-NOMA), hybrid successive interference cancellation (HSIC), power adaptation, energy efficiency.
\end{IEEEkeywords}

\section{Introduction}
As next-generation communication demands ultra-large-scale connectivity and high data rates, traditional orthogonal multiple access (OMA) technologies are becoming increasingly insufficient to meet these needs. Non-orthogonal multiple access (NOMA) techniques have received significant research interest in recent years due to their superior spectral efficiency and capability of realizing massive connectivity. By enabling multiple users to share identical resource blocks through power-domain multiplexing, NOMA achieves significant capacity gains compared to conventional orthogonal schemes \cite{dai2018survey,makki2020survey,you2021towards}. Furthermore, existing literature has demonstrated that NOMA maintains compatibility with advanced technologies, such as integrated sensing and communications (ISAC) \cite{sun2024study}, reconfigurable intelligent surface (RIS) networks \cite{li2023achievable,zhu2020power},  semantic communication \cite{mu2023exploiting}, {\color{black} and pinching antenna systems (PASS) \cite{Ding2025Pinching,Yang2025Pinching,Liu2025Pinching}. } 

However, since existing mobile networks are developed based on OMA, integrating NOMA into legacy frameworks to maximize its benefits while maintaining backward compatibility remains a critical technical challenge. Hybrid NOMA (H-NOMA), initially proposed for mobile edge computing (MEC) \cite{Ding2019MEC}, provides a feasible solution to integrate NOMA into legacy OMA-based systems \cite{liu2021latency,wei2022energy,Ding2025HNOMAopt,Fang2025Rethinking}. H-NOMA effectively combines the benefits of OMA and NOMA by allowing users to access both dedicated orthogonal resources and shared NOMA resources. This architecture ensures not only a seamless transition but also guaranteed backward compatibility.

In H-NOMA systems, each user divides its transmission into several sub-transmission phases, where
either OMA or NOMA will be adopted in each transmission phase. Therefore, it is evident that NOMA sub-transmission phases significantly impact the overall performance of H-NOMA systems.
Recall that a key challenge in NOMA is how to mitigate inter-user interference, for which successive interference cancellation (SIC) has emerged as an effective solution. The performance of NOMA systems is critically determined by the effectiveness of SIC decoding at the receiver.
Conventionally, fixed SIC (FSIC) methods employ predetermined decoding orders based on either channel state information (CSI) \cite{higuchi2013non,gao2017theoretical,Xia2018outage} or quality of service (QoS) criteria \cite{zhou2018state,Dhakal2019noma,ding2021new}. However, these approaches exhibit inherent limitations, most notably the persistent irreducible error floors in outage probability that occur even at high signal-to-noise ratios (SNRs), significantly compromising system reliability.
To overcome these limitations, hybrid SIC (HSIC) methods, which dynamically adjusts the decoding order of users by taking into account the relationship of several factors such as users's data rates, channel gains and transmit powers\cite{ding2021new,ding2020unveiling1,ding2020unveiling2,Yang2023HSICpower}, have been proposed and 
shown significant advantages on improving transmission robustness of pure NOMA.
Interestingly, the analytical results in \cite{Sun2025HSICnPA} show that, under certain conditions related to users’ target data rates, channel gains and power allocation ratios, the probability that the proposed H-NOMA scheme outperforms OMA in terms of data rate while consuming less energy approaches $1$, which indicates that H-NOMA generally provides better energy efficiency. Nevertheless, the required conditions for achieving this performance may not always hold in practical deployments. Thus, a natural and interesting question arises: whether the probability can approach zero without such constraints at high SNR, which motivates this paper.

Motivated by the aforementioned question, this paper proposes a novel H-NOMA uplink transmission scheme which is inspired by the recent developments on the application of power adaptation (PA) mechanisms to HSIC aided pure NOMA systems \cite{sun2021new,Sun2023Globecom}, 
termed ``HSIC-PA aided H-NOMA''. Rigorous analysis are provided for the proposed scheme, showing
significant performance improvement on transmission robustness. The main contributions of this paper are listed as follows: 
\begin{itemize}
\item A novel HSIC-PA aided H-NOMA scheme is proposed based on a TDMA legacy network. By applying power adaptation to the conventional HSIC aided H-NOMA scheme (hereafter referred to as HSIC-NPA aided H-NOMA scheme for the non-power adaptation version), the proposed approach achieves enhanced power efficiency and transmission robustness. A power adaptation coefficient $\gamma$ is introduced in opportunistic users to ensure that the proposed scheme maximizes the achievable data rate.
\item Rigorous derivations are conducted to obtain the closed-form expressions for the probability $\hat{P}_n$, i.e., the probability that the achievable rate of the proposed HSIC-PA-aided H-NOMA scheme underperforms OMA given a power reducing coefffcient $\beta<1/2$. The asymptotic analysis reveals that $\hat{P}_n \to 0$ in the high SNR regime regardless of channel conditions or power reducing coefffcient, implying deterministic superiority of the proposed scheme over both conventional OMA and HSIC-NPA aided H-NOMA.

\item Numerical results validate the theoretical analysis and demonstrate the superior performance of the HSIC-PA aided H-NOMA scheme. The results illustrate that the probability $\hat{P}_n$ converges to zero in high-SNR regime under all conditions, and it decays exponentially with the rate $n$, which means that the impact of $n$ (i.e., the order of the channel gain of the opportunistic user) on $\hat{P}_n$ is dominant compared to $m$ for both cases $m<n$ and $m>n$.

\end{itemize}

\section{System Model}
Consider an uplink communication scenario with one base station (BS) and $M$ users, each denoted by $U_i$, $1\le i\le M$. The users are ordered according to their channel gains with respect to the base station, $\left | h_{1} \right | ^2< \left | h_{2} \right | ^2<\cdots <\left |h_{M}\right|^2$, {\color{black} where $h_i$ denotes the normalized small scale Rayleigh fading coefficient for user $U_i$.}
Without loss of generality, time division multiple access (TDMA) is employed in a legacy OMA based network, where each user is allocated a distinct time slot of duration $T$. For simplicity, assume that the $i$-th time slot is assigned to user $U_i$ in every frame. The achievable data rate of user $U_i$ is given by $\log(1+\rho_i|h_i|^2)$,  where $\rho_i$ denotes the transmission power of $U_i$. It is noteworthy that the background noise of each user is normalized to $1$ in this paper. Meanwhile, the slow time-varying channel is considered, which means the channel gain of a user remains constant within a frame. 

In this paper, the H-NOMA scheme is considered to provide more opportunities to transmit for users. 
$M$ users are classified into two types in the considered H-NOMA scheme: legacy users and opportunistic users. 
Each legacy user can only transmit data within its assigned time slot as in the legacy OMA network. In contrast, each opportunistic user is allowed to transmit not only in its own time slot but also in the time slot of a legacy user by applying NOMA. In particular, the opportunistic user $U_n$ is grouped with the legacy user $U_m$, allowing $U_n$ to transmit its signal in both the $m$-th time slot (referred to as ``NOMA transmission'') and the $n$-th time slot (referred to as ``OMA transmission''). 

To demonstrate the advantages of the H-NOMA
scheme, the power consumption of $U_n$  must not exceed that of the benchmark OMA scheme. Consequently, the transmission power of $U_n$ is constrained to $\beta\rho_n$ in both the $m$-th and $n$-th slots, where $0<\beta < 1/2$ represents the power reduction coefficient. 
This constraint limits the transmission power of $U_n$ to  $\beta\rho_n$ in both NOMA and OMA time slots.

For NOMA transmission in the $m$-th slot, it is important to ensure that the transmission of $U_n$ does not degrade the transmission reliability of $U_m$ compared to the OMA scheme. The preset target rate of $U_m$ is assumed to be $R_m$, which represents the minimum rate required to ensure the successful transmission of $U_m$. The transmission of $U_n$ must be designed in a way that the outage probability of $U_m$ remains the same as in OMA. To address this requirement, three different NOMA schemes will be presented: the benchmark schemes are FSIC and HSIC-NPA aided H-NOMA schemes, as considered in previous work \cite{Sun2025HSICnPA}, and the last is the HSIC-PA aided H-NOMA scheme proposed in this paper. 

If $U_n$ works in traditional OMA mode, it transmits data only in the $n$-th time slot, with a transmission data rate 
\begin{equation}\label{equ_Rn}
    R_n = \log  \left (1+\rho_n \left| h_n \right|^2 \right ) 
\end{equation}

If $U_n$ works in the OMA transmission time slot of H-NOMA mode, i.e., the $n$-th time slot assigned for $U_n$, the data rate is obtained as 
\begin{equation}\label{equ_RnOMA}
    R^{\text{OMA}}_n = \log  \left (1+\beta\rho_n \left| h_n \right|^2 \right ) 
\end{equation}
Meanwhile, in the NOMA transmission time slot, i.e., the $m$-th time slot assigned for $U_m$, the data rate is different from each other since different NOMA schemes are applied, and will be discussed as follows.

If $U_n$ works in the NOMA transmission slot of FSIC aided H-NOMA scheme, as introduced in \cite{Sun2025HSICnPA}, the transmission rate of $U_n$ should be limited to  
\begin{equation}
\bar{R}^{\text{NOMA}}_n = \log \left(  1+  \frac{\beta \rho_n \left| h_n \right|^2  }{\rho_m \left| h_m \right|^2  +1} \right).
\end{equation}

If $U_n$ works in the NOMA transmission slot of HSIC-NPA aided H-NOMA scheme, as introduced in \cite{Sun2025HSICnPA}, the achievable data rate of $U_n$ is
\begin{align}
\!\!\tilde{R}^{\text{NOMA}}_{n} \!\!=\!\!\begin{cases}
 \tilde{R}^{I}_n \!=\!\log \!\left(\!1\!+\! {\beta \rho_n \left| h_n \right|^2  } \right)\!, \beta \rho_{n}\left | h_{n} \right|^2 \!\le\! \tau_m, \\  
\tilde{R}^{II}_n \!=\!\log \!\left(\! 1 \!+\!  \frac{\beta \rho_n \left| h_n \right|^2  }{\rho_m \left| h_m \right|^2  +1} \!\right)\!, \beta \rho_{n}\left | h_{n} \right |^2  \!>\! \tau_m, \!
\end{cases}
\end{align}
where $\tau_m = \max\left \{ 0,\frac{\rho_{ m}|h_m|^2 }{2^{R_{ m}} -1} -1\right \}$ is a threshold that represents the maximum interference level at which $U_m$ can still maintain the same outage performance as in the OMA scheme. Define  $\epsilon_m = 2^{R_m}-1$ and $\alpha_m = {\epsilon_m}/{\rho_m}$. The achievable rates $\tilde{R}^{I}_n$ and $\tilde{R}^{II}_n$ correspond to two distinct decoding orders for user $U_n$. These rates are determined by comparing the received power of $U_n$ at BS, i.e., $\beta \rho_n \left| h_n \right|^2$, with the interference threshold $\tau_m$, as shown in \cite{Sun2025HSICnPA}.

Although the HSIC-NPA aided H-NOMA scheme outperforms FSIC aided scheme, it still exhibits a critical limitation: the opportunistic user experiences severe interference when the legacy user's channel gain is sufficiently large. 
To address this limitation, an HSIC-PA aided H-NOMA scheme is proposed which adaptively decodes the opportunistic user's signal at either the first or second SIC stage by employing dynamic power adaptation. This approach selects the optimal decoding strategy to maximize the achievable rate as analyzed below.

\subsubsection{Type I}
The received power of $U_n$ at the BS is less than or equal to the interference threshold, i.e.,  $\beta \rho_n \left| h_n \right|^2 \leq \tau_m$. In this scenario, the signal of $U_m$ is decoded at the first stage of SIC same as HSIC-NPA Aided H-NOMA scheme, and the achievable rate of $U_n$ is 
\begin{equation}
    \hat{R}^I_n = \log \left(  1+  {\beta \rho_n \left| h_n \right|^2  } \right).
\end{equation}

\subsubsection{Type II}
The received power of $U_n$ at the BS is larger than the interference threshold, i.e., $\beta \rho_n \left| h_n \right|^2 > \tau_m$. There are two possible cases for data transmission and decoding order.
\begin{itemize}
\item Case 1: In this case,  the signal of $U_n$ is decoded at the first stage of SIC, and $U_n$'s data rate is the same as the FSIC one: 
\begin{equation}
    \hat{R}^{II,1}_n =  \log \left(  1+  \frac{\beta \rho_n \left| h_n \right|^2 }{\rho_m \left| h_m \right|^2  +1} \right).
\end{equation}
\item Case 2: In this case, power adaptation is introduced to provide an opportunity to achieve a higher data rate. The power adaptation factor $0< \gamma \leq 1$ is chosen such that $\gamma \beta \rho_n \left| h_n \right|^2 = \tau_m$.  As a result, the signal of $U_n$ can be decoded at the second stage of SIC, yielding the following achievable data rate of $U_n$:
\begin{equation}
    \hat{R}^{II,2}_n =  \log \left( 1+  \tau_m \right).
\end{equation}
\end{itemize}

Accordingly, the achievable rate of  $U_n$ in Type II can be expressed as
\begin{align}
      \hat{R}^{II}_n  
      = \max \left\{\hat{R}^{II,1}_n,\hat{R}^{II,2}_n \right\} = \log \left(\hat{r}^{II}_n\right).      
\end{align}
where 
\begin{align}   \label{equ_rnII}
\hat{r}^{II}_n = \max \left\{\left(1+  \tfrac{\beta \rho_n \left| h_n \right|^2}{\rho_m \left| h_m \right|^2  +1} \right), \left( 1+  \tau_m \right) \right\}.
\end{align}

Therefore, the achievable data rate of $U_n$ in NOMA slot of the HSIC-PA aided H-NOMA scheme is
\begin{align}
\hat{R}^{\text{NOMA}}_{n}= 
\begin{dcases}
 \hat{R}^{I}_n, 
 &\beta \rho_{n}\left | h_{n} \right|^2 \le   \tau_m \\  
\hat{R}^{II}_n,   &\beta \rho_{n}\left | h_{n} \right |^2  > \tau_m.
\end{dcases}
\end{align}

The probabilities that the achievable rates of the three considered H-NOMA schemes fail to surpass their OMA counterparts are are given by:
\begin{align}
\bar{P}_n &\!=\! \mathrm{Pr}\!\left(T \bar{R}^{\text{NOMA}}_n  \!+\! T R^{\text{OMA}}_n \leq T R_n\!\right), {\text{FSIC}},     \\
\tilde{P}_n &\!=\! \mathrm{Pr}\!\left(T \tilde{R}^{\text{NOMA}}_n \!+\! T R^{\text{OMA}}_n \leq T R_n\!\right), {\text{HSIC-NPA}} \\
\hat{P}_n &\!=\! \mathrm{Pr}\!\left(T \hat{R}^{\text{NOMA}}_n \!+\! T R^{\text{OMA}}_n \leq T R_n\!\right), {\text{HSIC-PA}}
\end{align}

Recall that the energy consumption for both FSIC aided H-NOMA and HSIC-NPA aided H-NOMA is $2T\beta\rho_n$, while for the HSIC-PA aided H-NOMA is $(1+\gamma)T\beta\rho_n$, which are all smaller than the OMA counterpart, since $0<\beta <1/2$ and $0<\gamma\leq1$. 


In \cite{Sun2025HSICnPA}, the probabilities of HSIC-NPA aided H-NOMA underperforming OMA in achievable rate $\tilde{P}_n$ are analyzed. Furthermore, the performance of FSIC aided and HSIC-NPA aided H-NOMA systems is compared, with numerical results highlighting the advantages of HSIC-NPA over FSIC. As demonstrated there, severe error floors occur for the probabilities $\bar{P}_n$ and $\tilde{P}_n$ at high SNRs under specific conditions. This indicates that both the FSIC aided and HSIC-NPA aided H-NOMA schemes fail to achieve further performance improvements even with significantly increased SNR.

\section{Performance Analysis for HSIC-PA Aided H-NOMA}
As shown in the previous section, three different H-NOMA schemes are introduced and their achievable data rates are derived. HSIC-PA has been proven to improve transmission performance for single time slot uplink NOMA \cite{sun2023hybrid}.
This paper demonstrates that HSIC-PA can further enhance the transmission efficiency of H-NOMA systems.  The probability of HSIC-PA aided H-NOMA achieving rates below the OMA baseline analyzed rigorously, providing key insights into its performance benefits.

In this section, exact closed-form expressions for $\hat{P}_n$ are derived for both $m < n$ and $m > n$ scenarios.
Based on these derivations,  asymptotic analyses are conducted in the high-SNR regime. Furthermore, detailed comparisons between HSIC-PA aided and HSIC-NPA aided H-NOMA schemes reveal substantial performance gains enabled by power adaptation.

The probability of OMA achieving superior performance compared to HSIC-PA aided H-NOMA can be expressed as \eqref{equ_HSIC_PA_Pn}.
\begin{figure*}
    \centering    
\begin{equation}    
\begin{split}
\label{equ_HSIC_PA_Pn}
\hat{P}_n 
= & \mathrm{Pr}\left(T \hat{R}^{\text{NOMA}}_n  + T R^{\text{OMA}}_n \leq T R_n\right) \\
= & \underbrace{P\left(\left(1+\beta\rho_n \left| h_n \right|^2 \right)\left(1+\beta\rho_n \left| h_n \right|^2 \right)  \leq  \left (1+\rho_n \left| h_n \right|^2 \right),\beta \rho_n \left| h_n \right|^2 \leq \tau_m \right)}_{P_{I}} \\
& +\underbrace{P\left(\hat{r}^{II}_n  \cdot \left(1+\beta\rho_n \left| h_n \right|^2 \right)  \leq \left (1+\rho_n \left| h_n \right|^2 \right),\beta \rho_n \left| h_n \right|^2 > \tau_m\right)}_{P_{II}} 
\end{split}
\end{equation}
\vspace{-1em}
\end{figure*}
The probability $P_I$ has been analyzed in (54) of \cite{Sun2025HSICnPA}, where it is denoted as $P_{1}$. 
Recall that $\tau_m$ takes two distinct values depending on the condition of $\left| h_m \right|^2 \alpha_m ^{-1}$. Consequently, $P_{II}$ can be divided into two distinct components, expressed as: 
\begin{equation}
    \begin{split}           
    \label{equ_P_II}
    {P_{II}} = &\underbrace{P\bigg(\hat{r}^{II}_n  \cdot \left(1+\beta\rho_n \left| h_n \right|^2 \right)  \leq \left (1+\rho_n \left| h_n \right|^2 \right),}_{P_{T}}  \\
    &\quad \underbrace{\beta \rho_n \left| h_n \right|^2 > \tau_m, \tau_m>0\bigg)}_{P_{T}} \\
     &+\underbrace{P\bigg(\hat{r}^{II}_n  \cdot \left(1+\beta\rho_n \left| h_n \right|^2 \right)  \leq \left (1+\rho_n \left| h_n \right|^2 \right),}_{P_{II,2}} \\
     &\quad\underbrace{\beta \rho_n \left| h_n \right|^2 > \tau_m, \tau_m=0\bigg)}_{P_{II,2}}.
    \end{split}
\end{equation}
Note that, when $\tau_m=0$ (i.e., when $\left| h_m \right|^2 <\alpha_m$), $\beta \rho_n \left| h_n \right|^2 > \tau_m$ holds universally, and $$\hat{r}^{II}_n \Big|_{\tau_m=0} = \left( 1+  \dfrac{\beta \rho_n \left| h_n \right|^2}{\rho_m \left| h_m \right|^2  +1} \right).$$
Accordingly, $P_{II,2}$ is denoted as
\begin{equation}
    \begin{split}
\label{equ_P_II2}
{P_{II,2}} 
=& P\bigg(\left| h_n \right|^2 \leq \frac{(1-\beta)(\rho_{m}\left|h_{m}\right|^{2}+1)-\beta}{\beta^{2}\rho_{n}}, \\
& \quad\left| h_m \right|^2 <\alpha_m \bigg).  
    \end{split}
\end{equation}
As analyzed in (56) of \cite{Sun2025HSICnPA}, this probability is denoted as $P_{2,2}$. It is important to observe the following lemma, which divides $P_T$ into two parts.
\begin{lemma}
The probability $P_T$ can be expressed as the sum of two probabilities, as shown in \eqref{equ_PT}. 
\begin{figure*} 
\centering
\begin{equation}
\begin{split}
    \label{equ_PT}
{P_{T}} = & \underbrace{P(\left|h_n\right|^2>\Theta(\left|h_{m}\right|^{2}),
\left|h_n\right|^2<\Omega(\left|h_{m}\right|^{2}),
\left|h_n\right|^2>\Phi(\left|h_{m}\right|^{2}),
\left|h_m\right|^2>\alpha_m)}_{P_{T,1}} \\
& + 
\underbrace{P(\left|h_{n}\right|^{2}<\Psi(\left|h_{m}\right|^{2}),
\left|h_{n}\right|^{2}>\Omega(\left|h_{m}\right|^{2}),
\left|h_{n}\right|^{2}>\Phi(\left|h_{m}\right|^{2}),
\left|h_{m}\right|^{2}>\alpha_{m})}_{P_{T,2}}
\end{split}    
\end{equation}
\vspace{-1em}
\end{figure*}
The variables in \eqref{equ_PT} are defined as follows:
\begin{equation}
\begin{split}   
\Phi(\left|h_{m}\right|^{2})=&\frac{\left|h_{m}\right|^{2}\alpha_{m}^{-1}-1}{\beta\rho_{n}}, \\
\Omega(\left|h_{m}\right|^{2})=&\frac{(\left|h_{m}\right|^{2}\alpha_{m}^{-1}-1)(1+\rho_{m}\left|h_{m}\right|^{2})}{\beta\rho_{n}}, \\
\Theta(\left|h_{m}\right|^{2})=&\frac{\left|h_{m}\right|^{2}\alpha_{m}^{-1}-1}{1-\beta\left|h_{m}\right|^{2}\alpha_{m}^{-1}}\cdot\frac{1}{\rho_{n}}, \\
\Psi(\left|h_{m}\right|^{2})=&\frac{(1-\beta)(\rho_{m}\left|h_{m}\right|^{2}+1)-\beta}{\beta^{2}\rho_{n}}.
\end{split}
\end{equation}
\begin{IEEEproof}
Please refer to Appendix A.
\end{IEEEproof}
\end{lemma}

Based on Lemma $1$, the expressions for $\hat{P}_n$ in the scenarios where $m<n$ and $m>n$ can be obtained, which are shown in the following two subsections, respectively.
\subsection{$m<n$}
\begin{theorem}
When $m<n$,  $\hat{P}_n$  can be expressed as
\begin{align}
\hat{P}_n=P_1+P_T+P_{2,2},
\end{align}
where $P_T=P_{T1,1}+P_{T1,2}+P_{T1,3}+P_{T2,1}+P_{T2,2}$.  
$P_1$ and $P_{2,2}$ have been presented in \cite{Sun2025HSICnPA}, while $P_{T1,1}$, $P_{T1,2}$, $P_{T1,3}$, $P_{T2,1}$ and $P_{T2,2}$ are summarized in Table I and Table II, respectively. 
The expressions of the variables are listed as follows:
\begin{equation}
\begin{split}
k_{1}&=\frac{1-2\beta}{(1-\beta)\beta\varepsilon_m},\\
k_{2}&=\frac{1-\beta}{\beta^2} + \frac{1-2\beta}{\beta^2 \varepsilon_m},\\
k_{3}&=\frac{1-2\beta}{(1-\beta)\beta\varepsilon_m} + \frac{1-2\beta}{\beta^2}.
\end{split}
\end{equation}

\begin{table*}[htbp!]
\caption{The expressions for $P_1$, $P_{T1,1}$, $P_{T1,2}$ and $P_{T1,3}$, when $m<n$.}
    \begin{center}
    \renewcommand{\arraystretch}{1.5}
        \begin{tabular}{|c|c|c|c|c|c|}
        \hline
        & & $\frac{\rho_n}{\rho_m} \le k_{1}$ & $k_{1} < \frac{\rho_n}{\rho_m} \le\frac{1-\beta}{\beta\varepsilon_m}$ & $\frac{1-\beta}{\beta\varepsilon_m} < \frac{\rho_n}{\rho_m} \le \frac{1}{\beta\varepsilon_m}$ & $\frac{\rho_n}{\rho_m} > \frac{1}{\beta\varepsilon_m}$ \\
        \hline
        \multirow{3}{*}{Exact results} & $P_1$ \cite{Sun2025HSICnPA} & $T_1$ & 0 & 0 & 0 \\  \cline{2-6}
        & $P_{T1,1}$ & $S_1$ & 0 & 0 & 0 \\  \cline{2-6}
         & $P_{T1,2}$ & $S_2$ & $S_3$ & $S_3$ & $S_3$ \\  \cline{2-6}
         & $P_{T1,3}$ & $S_4$ & $S_5$ & $S_5$ & $S_5$ \\  \cline{2-6}
        \hline
        \multirow{3}{*}{Approximations}& $P_1$ \cite{Sun2025HSICnPA}& $\frac{1}{\rho_m^n} \tilde{T}_1$ & 0 & 0 & 0 \\  \cline{2-6} 
        & $P_{T1,1}$ & $\frac{1}{\rho_m^n} \tilde{S}_1$ & 0 & 0 & 0 \\  \cline{2-6}
         & $P_{T1,2}$ & $\frac{1}{\rho_m^n} \tilde{S}_2$ & $\frac{1}{\rho_m^n} \tilde{S}_3$ & $\frac{1}{\rho_m^n} \tilde{S}_3$ & $  \frac{1}{\rho_m^n} \tilde{S}_3$ \\  \cline{2-6}
         & $P_{T1,3}$ & $\frac{1}{\rho_m^n} \tilde{S}_4$ & $\frac{1}{\rho_m^n} \tilde{S}_5$ & $\frac{1}{\rho_m^n} \tilde{S}_5$ & $\frac{1}{\rho_m^n} \tilde{S}_5$ \\  \cline{2-6}
        \hline
        \end{tabular}
    \end{center}
\label{tab1_m<n_P1}
\end{table*}

\begin{table*}[htbp!]
\caption{The expressions for $P_{T2,1}$, $P_{T2,2}$ and $P_{2,2}$, when $m<n$.}
    \begin{center}
    \renewcommand{\arraystretch}{1.5}
        \begin{tabular}{|c|c|c|c|c|}
        \hline
         && $\frac{\rho_n}{\rho_m} \le \frac{1-\beta}{\beta^2}$ & $\frac{1-\beta}{\beta^2} < \frac{\rho_n}{\rho_m} \le k_{2}$ & $\frac{\rho_n}{\rho_m} > k_{2}$ \\
        \hline
        \multirow{2}{*}{Exact results} 
        & $P_{T2,1}$ & $S_6$                 & $S_7$  & 0                  \\  \cline{2-5}
        & $P_{T2,2}$ & $S_8$    & $S_8$    & $S_8$ \\  \cline{2-5}
        & $P_{2,2}$ {\cite{Sun2025HSICnPA}} & $T_6$    & $T_6$    & $T_7$ \\  \cline{2-5}
        \hline
        \multirow{2}{*}{Approximations} 
        & $P_{T2,1}$ & $\frac{1}{\rho_m^n} \tilde{S}_6$              & $\frac{1}{\rho_m^n} \tilde{S}_7$ 
        & 0 \\   \cline{2-5}
        & $P_{T2,2}$ & $\frac{1}{\rho_m^n} \tilde{S}_8$  & $\frac{1}{\rho_m^n} \tilde{S}_8$                       
        & $\frac{1}{\rho_m^n} \tilde{S}_8$ \\  \cline{2-5}
        & $P_{2,2}${\cite{Sun2025HSICnPA}} & $\frac{1}{\rho_m^n} \tilde{T}_6$  & $\frac{1}{\rho_m^n} \tilde{T}_6$         & $\frac{1}{\rho_m^n} \tilde{T}_7$ \\  \cline{2-5}
        \hline
        \end{tabular}
    \end{center}
\label{tab2_m<n_P2}
\end{table*}

\begin{align}
    \begin{split}  
    S_{1}={c_{mn}}\!\!\!\sum\limits_{p = 0}^{n - m - 1} \!\!\!{{c_p}\sum\limits_{l = 0}^{m - 1} {{c_l}} }\Big[e^{\frac{M-m-p}{\beta\rho_{n}}}\frac{e^{-r_1\omega_1}-e^{-r_1\omega_2}}{r_1}\\-e^{\frac{M-m-p}{\beta\rho_{n}}}\Gamma_{1}(\omega_1,\omega_2,A,B)\Big]
    \end{split}
\end{align}
where $\omega_1=\frac{1}{\alpha_m^{-1}-\beta\rho_n}$, $\omega_2=\frac{1-\beta}{\beta}\alpha_m$,  $c_l=\Big(\!\!\begin{array}{c} {m-1}\\l \end{array}\!\!\Big)(-1)^l$, $c_p=\Big(\!\!\begin{array}{c} {n-m-1}\\{l} \end{array}\!\!\Big)(-1)^{n-m-1-p}$, $r_1=l+p+1+\frac{M-m-p}{\beta\rho_n\alpha_m}$, $c_{mn}=\frac{M!}{(m-1)!(n-m-1)!(M-n)!}$,  
\begin{align}   
\begin{split}    
\Gamma_{1}(a,b,c,d)=&\frac{e^{\frac{d^{2}}{4c}}\sqrt{\pi}}{2\sqrt{c}}\bigg(erf\Big(\sqrt{c}\big(b+\frac{d}{2c}\big)\Big) \\
&-erf\Big(\sqrt{c}\big(a+\frac{d}{2c}\big)\Big)\bigg),
\end{split}
\end{align}
$A=\frac{(M-m-p)\rho_m\alpha_m^{-1}}{\beta\rho_n}$, $B=\frac{M-m-p}{\beta\rho_n}\left(\alpha_m^{-1}-\rho_m\right)+l+p+1$, and
$erf\left(x\right)=\frac{2}{\sqrt{\pi}}\int_{0}^{x}e^{-t^{2}}dt$.
\begin{align}
    S_2 &=W_1(\frac{z_1-\omega_2}{2},\frac{z_1+\omega_2}{2}),  \\
    S_3 &=  \begin{dcases}
    0,&z_2>z_1\\
  W_1(\frac{z_1-z_2}{2},\frac{z_1+z_2}{2}),&z_2<z_1      
    \end{dcases}
\end{align}
where ${z}_{1}=\frac{\alpha_{m}}{2\beta}-\frac{1}{2\rho_{m}}+\frac{1}{2}\sqrt{(\frac{\alpha_{m}}{\beta}-\frac{1}{\rho_{m}})^{2}+4\frac{(1-\beta)}{\beta\rho_{m}}}$, $z_{2}=\frac{\alpha_{m}}{2\beta}-\frac{1}{2\beta\rho_{n}}+\frac{1}{2}\sqrt{(\frac{\alpha_{m}}{\beta}-\frac{1}{\beta\rho_{n}})^{2}+4\frac{\alpha_{m}}{\beta\rho_{n}}}$,  
\begin{align}
\begin{split}
    W_1(a,b)=&{c_{mn}}\!\!\!\sum\limits_{p = 0}^{n - m - 1} \!\!\!{{c_p}\sum\limits_{l = 0}^{m - 1} {{c_l}} } \frac{a}{{M - m - p}}\\
    &\times\frac{\pi }{{{n_c}}}\sum\limits_{i = 1}^{{n_c}} \Gamma_2(at_{i}+b)\sqrt{1-{t_{i}}^{2}},     
\end{split}
\end{align}    
\begin{align}  
\begin{split}
\Gamma_2(x)=e^{-(l+p+1)x}\Big(&e^{\frac{(M-m-p)}{\rho_{n}}\frac{\alpha_{m}^{-1}x-1}{1-\beta\alpha_{m}^{-1}x}}\\
&-e^{-(M-m-p)\frac{(\alpha_{m}^{-1}x-1)(1+\rho_{m}x)}{\beta\rho_{n}}}\Big),
\end{split}
\end{align}
$n_c$ is parameter for Gauss-Chebyshev approximation as shown in \eqref{equ_GaussCheb}, and $t_{i}=\cos(\frac{2i-1}{2n_{c}}\pi)$.
\begin{align}
    {S_4}=W_2(z_3,\omega_1,A,B), 
    {S_5}=
    \begin{dcases}
    0,&z_2<z_3\\
          W_2(z_3,z_2,A,B),&z_2>z_3  
    \end{dcases}
\end{align}
where 
\begin{align}
    z_3 \!=\!\frac{1}{2}(\beta\eta\alpha_{m}\!+\!\alpha_{m}\!-\!\frac{1}{\rho_{m}}\!+\!\sqrt{(\beta\eta\alpha_{m}\!+\!\alpha_{m}\!-\!\frac{1}{\rho_{m}})\!+\!\frac{4\alpha_{m}}{\rho_{m}}})    
\end{align}
\begin{align} 
\begin{split}    
W_2(a,b,c,d)=&c_{mn}\!\!\!\sum_{p=0}^{n-m-1}\!\!\!c_{p}\sum_{l=0}^{m-1}c_{l}\frac{1}{M-m-p}\\
&\times\left[\Gamma_3(a,b)-e^{\frac{M-m-p}{\beta\rho_{n}}}\Gamma_{1}(a,b,c,d)\right],
\end{split}
\end{align}
\begin{align}    
\Gamma_3(a,b)=\frac{e^{-(M-m+l+1)a}-e^{-(M-m+l+1)b}}{M-m+l+1}.
\end{align}
\begin{align}
    S_6=W_3(\alpha_m,z_3), \quad
    {S_7}=
\begin{dcases}
 W_3(\alpha_m,z_3), &z_3<\omega_4\\
 W_3(\alpha_m,\omega_4),&z_3>\omega_4   
\end{dcases}
\end{align}
where 
\begin{align}
\begin{split}    
W_3(a,b)&={c_{mn}}\!\!\!\sum\limits_{p = 0}^{n - m - 1} \!\!\!{{c_p}\sum\limits_{l = 0}^{m - 1} {c_l} } \frac{1}{{M - m - p}} \\ 
 \times &\left[\Gamma_3(a,b)-{e^{ - \left( {M - m - p} \right){\omega _3}}}\frac{{{e^{ - r_2{a}}} - {e^{ - r_2{b}}}}}{r_2} \right],
\end{split}
\end{align}
$r_2=l+p+1+\frac{(M-m-p)(1-\beta)\rho_m}{\beta^2\rho_n}$, $\omega_3=\frac{1-2\beta}{\beta^2\rho_n}$, and $\omega_4=\frac{1-2\beta}{\beta^2\rho_n-(1-\beta)\rho_m}$. 
\begin{align}
     {S_8}=&
     \begin{dcases}
     0,&z_3>z_1\\
     W_4(z_3,z_1,A,B),&z_3<z_1
\end{dcases}
\end{align}
\begin{align}
\begin{split}
 W_4(a,b,c,d)=&c_{mn}\!\!\!\sum_{p=0}^{n-m-1}\!\!\!c_{p}\sum_{l=0}^{m-1}c_{l}\frac{1}{M-m-p} \\
 \times\bigg[&e^{\frac{M-m-p}{\beta\rho_{n}}}\Gamma_{1}(a,b,c,d) \\
 -&e^{-(M-m-p)\omega_{3}}\frac{e^{-r_2 a}-e^{-r_2 b}}{r_2}\bigg]     
\end{split}
\end{align}

\begin{IEEEproof}
	Please refer to Appendix B.	
\end{IEEEproof}
\end{theorem}

By applying Taylor expansions, the extreme value of $\hat{P}_n$ can be obtained in the high SNR regime. Since the extreme values of $P_1$ and $P_{2,2}$ have been described in \cite{Sun2025HSICnPA}, only the extreme value of $P_T$ when $m>n$ is presented here, as highlighted in the following corollary.
\begin{corollary}
For the case $m<n$, when $\rho_n \to \infty$, $\rho_m \to \infty$, and $\frac{\rho_n}{\rho_m}=\eta$ is a constant, the value of $P_T$ can be approximated as summarized in Table I  and Table II, respectively.
The expressions for the variables used in Table I and Table II can be expressed as follows:
\begin{align}
\begin{split}
    \tilde{S}_1=&c_{mn}\!\!\!\sum_{p=0}^{n-m-1}\!\!\!\binom{n-m-1}{p}\frac{(-1)^p}{n-m-p}\Bigg[\Gamma_4(\varpi_1,\varpi_2)\\
    -&\sum_{q=0}^{n-m-p}\frac{(\frac{1}{\varepsilon_m})^{n-m-p-q}(-1)^q({\varpi_2}^{n-q}-{\varpi_1}^{n-q})}{(\beta\eta)^{n-m-p}(n-q)}\Bigg]
\end{split}
\end{align}
where $\varpi_1=\frac{1}{\epsilon_m^{-1}-\beta\eta}$, $\varpi_2=\frac{1-\beta}{\beta}\varepsilon_{m}$, $\eta = \frac{\rho_n}{\rho_m}$, and
\begin{align}
     \label{equ_Gamma3}
     \Gamma_4(a,b)\!=\!\!&\sum\limits_{i_{1}+i_{2}+i_{3}=n-m-p}\!\!\!\binom{n\!-\!m\!-\!p}{i_{1},i_{2},i_{3}}\left(\!\frac{1}{\varepsilon_{m}}\!\right)^{i_1}\left(\frac{1}{\varepsilon_{m}}\!-\!1\right)^{i_{2}}  \nonumber\\
     \times &(-1)^{i_{3}}\frac{\left(b^{m+p+2i_{1}+i_{2}}-a^{m+p+2i_{1}+i_{2}}\right)}{\left(\beta\eta)^{n-m-p}(m+p+2i_{1}+i_{2}\right)}. 
\end{align}
\begin{align}
    \tilde{S}_2=W_5(\varpi_{2},\bar{z}_{1}), \quad \tilde{S}_3=
    \begin{dcases}
    0,&z_1<z_2\\
      W_5(\bar{z}_{2},\bar{z}_{1}),&z_1>z_2 
    \end{dcases}
\end{align}
where $\overline{z}_{1}=\frac{\varepsilon_{m}}{\beta}-1+\sqrt{\left(\frac{\varepsilon_{m}}{\beta}-1\right)^{2}+4\frac{\left(1-\beta\right)}{\beta}\varepsilon_{m}}$,  $\overline{z}_{2}=\frac{\varepsilon_{m}}{\beta}-\frac{1}{\beta\eta}+\sqrt{\left(\frac{\varepsilon_{m}}{\beta}-\frac{1}{\beta\eta}\right)^{2}+4\frac{\varepsilon_{m}}\beta\eta}$, and
\begin{align}
\begin{split}
W_5(a,b)=&c_{mn}\sum_{p=0}^{n-m-1}\binom{n-m-1}{p}\frac{\left(-1\right)^{p}}{n-m-p} \\
&\times \left[\Gamma_4(a,b)-\Gamma_5(a,b)\right],
\end{split}
\end{align}

\begin{align}
\label{equ_Gamma4}
\Gamma_5(a,b)=&\frac{1}{\eta^{n-m-p}}\sum\limits_{q+j=0}^{\infty} \sum\limits_{q=0}^{\min (q+j, n-m-p)}\!\!\!\binom{n-m-p}{q} \nonumber\\
&\times\binom{n-m-p+j-1}{j} (-1)^{n-m-p-q}\\
&\times \frac{\beta^{j}}{\varepsilon_{m}^{q+j}}(b^{m+p+q+j}-a^{m+p+q+j}). \nonumber
\end{align}
\begin{align} 
\tilde{S}_4 = W_6(\overline{z}_{3},\varpi_{1}), \quad
\tilde{S}_5 =
\begin{dcases}
0,&z_3>z_2\\
W_6(\overline{z}_{3},\overline{z}_{2}), &z_3<z_2   
\end{dcases}
\end{align}
where $\overline{z}_{3}\!=\!\beta\eta\varepsilon_{m}\!+\!\varepsilon_{m}\!-\!1 \!+\!\sqrt{(\beta\eta\varepsilon_{m}\!+\!\varepsilon_{m}\!-\!1)\!+\!4\varepsilon_{m}}$, and 
\begin{align} 
\begin{split}
W_6(a,b)=&c_{mn}\!\!\!\sum_{p=0}^{n-m-1}\!\!\!\binom{n-m-1}{p}\frac{\left(-1\right)^{p}}{n-m-p}\\
&\times \left[\Gamma_4\left(a,b\right)-\frac{b^{n}-a^{n}}{n}\right].
\end{split}
\end{align}
\begin{align}
    \tilde{S}_6=W_7(\varepsilon_{m},\overline{z}_{3}), \quad
    \tilde{S}_7=
    \begin{dcases}
    W_7(\varepsilon_{m},\overline{z}_{3}), &z_3<\omega_4\\ W_7(\varepsilon_{m},\varpi_{4}),&z_3>\omega_4
    \end{dcases}
\end{align}
where $\varpi_4=\frac{1-2\beta}{\beta^2\eta-(1-\beta)}$, and
\begin{align}
\begin{split}
    W_7(a,b)=&c_{mn}\!\!\!\sum_{p=0}^{n-m-1}\!\!\!\binom{n-m-1}{p}\frac{\left(-1\right)^{p}}{n-m-p}\\
    &\times \left[\Gamma_6(a,b)-\frac{b^{n}-a^{n}}{n}\right].
\end{split}
\end{align}
\begin{align}
\begin{split}    
\Gamma_6(a,b)=&\sum\limits_{q=0}^{n-m-p}\binom{n-m-p}{q}\\
\times&\frac{(1-\beta)^{n-m-p-q}(1-2\beta)^q(b^{n-q}-a^{n-q})}{(\beta^2\eta)^{n-m-p}(n-q)}, 
\end{split}
\end{align}
\begin{align}
    \tilde{S}_8=&
    \begin{dcases}
    0,&z_3>z_1\\
    W_8(\overline{z}_{3},\overline{z}_{1}),&z_3<z_1
    \end{dcases}
\end{align}
\begin{align}
\begin{split}    
    W_8(a,b)=& c_{_{mn}}\!\!\!\sum_{p=0}^{n-m-1}\!\!\!\binom{n-m-1}{p}\frac{(-1)^{^p}}{n-m-p}\\
    &\times \left[\Gamma_6(a,b)-\Gamma_4(a,b)\right] 
\end{split}
\end{align}

\begin{IEEEproof}
	Please refer to Appendix C.	
\end{IEEEproof}
\end{corollary}

Based on the results shown in Corollary $1$, interesting insights can be obtained as highlighted in the following remarks.


{\color{black} 
\begin{Remark}
From the results in Table I and Table II for the case $m<n$, it can be observed that the probability ${P}_T$ in the proposed H-NOMA scheme approaches 0 as $\rho_n$ and $\rho_m$ tend to infinity under all conditions. In contrast, in the HSIC-NPA aided H-NOMA scheme, ${P}_{2,1}$ which is shown in Table I of \cite{Sun2025HSICnPA} does not approach 0 when both $\epsilon_m>\frac{\beta}{1-\beta}$ and $\frac{\rho_n}{\rho_m}\le \frac{1-\beta}{\beta^2}$ hold. 
\end{Remark}

\begin{Remark}
Since ${P}_{2,1}$ in Table I of \cite{Sun2025HSICnPA} is the inherent cause of the error floor in $\tilde{P}_n$ for HSIC-NPA aided H-NOMA scheme under certain constraints, the proposed method effectively resolves this issue. Consequently, the probability $\hat{P}_n$ approaches 0 in high SNR regime under all conditions for the case $m<n$, demonstrating the performance superiority of the proposed scheme.
\end{Remark}
}

\begin{Remark}
For the case $m<n$, when $\rho_n \to \infty$, $\rho_m \to \infty$, and $\frac{\rho_n}{\rho_m}$ is a constant, it can be observed that $\hat{P}_n$ decays exponentially with the rate $n$, i.e., $\hat{P}_n\propto \frac{1}{\rho_m^n}$ or $\hat{P}_n\propto \frac{1}{\rho_n^n}$, 
which means that the impact of $n$ (i.e., the order of the channel gain of the opportunistic user) on $\hat{P}_n$ is dominant comparing to $m$.
\end{Remark}

\subsection{$m>n$}
\begin{theorem}
When $m>n$,  $\hat{P}_n$  can be expressed as
\begin{align}
\hat{P}_n=P_1+P_T+P_{2,2},
\end{align}
where $P_T=P'_{T1,1}+P'_{T1,2}+P'_{T1,3}+P'_{T1,4}+P'_{T2,1}+P'_{T2,2}$. $P_1$ and $P_{2,2}$ have been presented in \cite{Sun2025HSICnPA}, while $P'_{T1,1}$, $P'_{T1,2}$, $P'_{T1,3}$, $P'_{T1,4}$, $P'_{T2,1}$ and $P'_{T2,2}$ are summarized in Table III and Table IV, respectively.
The expressions of the variables are listed as follows:
\begin{align}
    V_1=H_1(\alpha_{m},z_{3},D,E), \quad V_2=H_1(\alpha_{m},\omega_{2},D,E),
\end{align}
where 
\begin{align}    
\begin{split}    
H_1(a,b,c,d)=&\hat{c}_{mn}\!\!\!\sum_{p=0}^{m-n-1}\!\!\!\hat{c}_{p}\sum_{l=0}^{n-1}\hat{c}_{l}\frac{1}{l+p+1}e^{\frac{l+p+1}{\beta\rho_{n}}} \\
&\times \left[\frac{e^{-r_3a}-e^{-r_3b}}{r_3}-\Gamma_1(a,b,c,d)\right],
\end{split}
\end{align}
$\hat{c}_{mn}=\frac{M!}{(n-1)!(m-n-1)!(M-m)!}$, $\hat{c}_l=\left(\!\!\begin{array}{c} {n-1}\\{l}\end{array}\!\!\right)(-1)^l$, $\hat{c}_p=\left(\!\!\begin{array}{c} {m-n-1}\\{l} \end{array}\!\!\right)(-1)^{m-n-1-p}$, $r_3=M-n-p+\frac{l+p+1}{\beta\rho_n\alpha_m}$, $D=\frac{\rho_m\alpha_m^{-1}}{\beta\rho_n}(l+p+1)$, and $E=\frac{(l+p+1)(\alpha_m^{-1}-\rho_m)}{\beta\rho_n}+M-n-p$.
\begin{align}
    V_3=H_2(z_3,\omega_1), \quad V_4=H_2(z_3,\omega_2),
\end{align}
where 
\begin{align}
\begin{split}
    H_2(a,b)=&{{\hat c}_{mn}}\!\!\!\sum\limits_{p = 0}^{m - n - 1}\!\!\! {{{\hat c}_p}} \sum\limits_{l = 0}^{n - 1} {{{\hat c}_l}} \frac{1}{{l + p + 1}}\\
    &\times\left[ {{e^{\frac{{l + p + 1}}{{\beta {\rho _n}}}}}\frac{{{e^{ - r_3a}} - {e^{ - r_3{b}}}}}{r_3} - \Gamma_7(a,b)} \right],
\end{split}
\end{align}
\begin{align}
    \Gamma_7(a,b)=&\frac{e^{-(M-n+l+1)a}-e^{-(M-n+l+1)b}}{M-n+l+1}.
\end{align}
\begin{align}
    V_5=&
    \begin{dcases}
    H_3(\omega_2,z_1),&z_1<z_3\\
   H_3(\omega_2,z_3),&z_1>z_3
    \end{dcases}
\end{align}
where 
\begin{align}
\begin{split}
    H_3(a,b)=&{{\hat c}_{mn}}\!\!\!\sum\limits_{p = 0}^{m - n - 1} \!\!{{{\hat c}_p}} \sum\limits_{l = 0}^{n - 1} {{{\hat c}_l}} \frac{1}{{l + p + 1}}\frac{{{b} - {a}}}{2}\\
    &\times \frac{\pi }{{{n_c}}}\sum\limits_{i = 1}^{{n_c}} \Gamma_{8}(\frac{b-a}{2}t_{i}+\frac{a+b}{2})\sqrt{1-t_{i}^{2}}, 
\end{split}
\end{align}
\begin{align}
\begin{split}
    \Gamma_{8}(x)=&e^{-(M-n-p)x}\Big(e^{-\frac{(l+p+1)}{\rho_{n}}\frac{\alpha_{m}^{-1}x-1}{1-\beta\alpha_{m}^{-1}x}}\\
    &-e^{-(l+p+1)\frac{(\alpha_{m}^{-1}x-1)(1+\rho_{m}x)}{\beta\rho_{n}}}\Big).
\end{split}
\end{align}
\begin{align}
    V_6=H_4(\omega_2,z_2),\quad
    V_7=
    \begin{dcases}
        0,&z_2<z_3\\
           H_4(z_3,z_2),&z_2>z_3 
    \end{dcases}
\end{align}
where 
\begin{align}
\begin{split}    
    H_4(a,b)=&{{\hat c}_{mn}}\sum\limits_{p = 0}^{m - n - 1} {{{\hat c}_p}} \sum\limits_{l = 0}^{n - 1} {{{\hat c}_l}} \frac{1}{{l + p + 1}}\Bigg[\frac{{{b} - {a}}}{2}\frac{\pi }{{{n_c}}} \\
 \times\Big(\sum\limits_{i = 1}^{{n_c}}\Gamma_{9}&\big(\frac{b-a}{2}t_{i}+\frac{b+a}{2}\big)\sqrt{1-t_{i}^{2}}\Big)- \Gamma_7(a,b)\Bigg],
\end{split}
\end{align}
\begin{align}
\Gamma_{9}(x)=&e^{-(M-n-p)x}e^{-\frac{(l+p+1)}{\rho_{n}}\frac{\alpha_{m}^{-1}x-1}{1-\beta\alpha_{m}^{-1}x}}.
\end{align}
\begin{align}
    V_8=H_5(\alpha_m,z_3,D,E), 
\end{align}
\begin{align}
    V_{9}=
    \begin{dcases}
    H_5(\alpha_m,z_3,D,E),&z_3<\omega_4\\
    H_5(\alpha_m,\omega_4,D,E),&z_3>\omega_4
    \end{dcases}
\end{align}
where 
\begin{align}
\begin{split}
    H_5(a,b,c,d)=&\hat{c}_{mn}\sum_{p=0}^{m-n-1}\hat{c}_{p}\sum_{l=0}^{n-1}\hat{c}_{l}\frac{1}{l+p+1}\\
    &\times\Bigg[e^{\frac{l+p+1}{\beta\rho_{n}}}\Gamma_1(a,b,c,d)-\Gamma_7(a,b)\Bigg].
\end{split}
\end{align}
\begin{align}
    V_{10}=
    \begin{dcases}
    0,&\omega_4>z_1\\
    H_6(\omega_{4},z_{1},D,E),&\omega_4<z_1
    \end{dcases},
\end{align}
\begin{align}
    V_{11}=H_6(\alpha_{m},z_{1},D,E)
\end{align}
where 
\begin{align}
\begin{split}    
    H_6(a,b,c,d)=&\hat{c}_{mm}\!\!\!\sum_{p=0}^{m-n-1}\!\!\!\hat{c}_{p}\sum_{l=0}^{n-1}\hat{c}_{l}\frac{1}{l+p+1} \\ 
    &\times\bigg[e^{\frac{l+p+1}{\beta\rho_{n}}}\Gamma_1(a,b,c,d)\\
    &-e^{-(l+p+1)\omega_{3}}\frac{e^{-r_4a}-e^{-r_4b}}{r_4}\bigg],    
\end{split}
\end{align}
and $r_4=M-n-p+\frac{(l+p+1)(1-\beta)\rho_m}{\beta^2\rho_n}$.
\begin{table*}[htbp!]
\caption{The expressions for  $P_1$, $P'_{T1,1}$, $P'_{T1,2}$, $P'_{T1,3}$ and $P'_{T1,4}$, when $m>n$.}
    \begin{center}
    \renewcommand{\arraystretch}{1.5}
        \begin{tabular}{|c|c|c|c|c|}
        \hline
         && $\frac{\rho_n}{\rho_m} \le k_{1}$ & $k_{1} < \frac{\rho_n}{\rho_m} \le k_{3}$ & $\frac{\rho_n}{\rho_m} > k_{3}$ \\  
        \hline
        \multirow{4}{*}{Exact results} & $P_1$ \cite{Sun2025HSICnPA} & $Q_1$ & $Q_2$ & $Q_2$ \\  \cline{2-5}
        & $P'_{T1,1}$ & $V_1$ & $V_1$ & $V_2$ \\  \cline{2-5}
         & $P'_{T1,2}$ & $V_3$ & $V_4$ & 0 \\  \cline{2-5}
         & $P'_{T1,3}$ & 0 & 0 & $V_5$ \\  \cline{2-5}
         & $P'_{T1,4}$ & 0 & $V_6$ & $V_7$ \\  \cline{2-5}
        \hline
        \multirow{4}{*}{Approximations}& $P_1$ \cite{Sun2025HSICnPA}& $\frac{1}{\rho_m^n}\tilde{S}+\frac{1}{\rho_m^m} \tilde{Q}_1$ & $\frac{1}{\rho_m^n}\tilde{S}+\frac{1}{\rho_m^m} \tilde{Q}_2$ & $\frac{1}{\rho_m^n}\tilde{S}+\frac{1}{\rho_m^m} \tilde{Q}_2$ \\  \cline{2-5}
        & $P'_{T1,1}$ & $\frac{1}{\rho_m^m} \tilde{V}_1$ & $\frac{1}{\rho_m^m} \tilde{V}_1$ & $\frac{1}{\rho_m^m} \tilde{V}_2$ \\  \cline{2-5}
         & $P'_{T1,2}$ & $\frac{1}{\rho_m^m} \tilde{V}_3$ & $\frac{1}{\rho_m^m} \tilde{V}_4$ & 0 \\  \cline{2-5}
         & $P'_{T1,3}$ & 0 & 0 & $\frac{1}{\rho_m^m} \tilde{V}_5$ \\  \cline{2-5}
         & $P'_{T1,4}$ & 0 & $\frac{1}{\rho_m^m} \tilde{V}_6$ & $\frac{1}{\rho_m^m} \tilde{V}_7$ \\  \cline{2-5}
        \hline
        \end{tabular}
    \end{center}
\label{tab3_m>n_P1}
\end{table*}

\begin{table*}[htbp!]
\caption{The expressions for $P'_{T2,1}$, $P'_{T2,2}$ and $P_{2,2}$, when $m>n$.}
    \begin{center}
    \renewcommand{\arraystretch}{1.5}
        \begin{tabular}{|c|c|c|c|c|}
        \hline
         && $\frac{\rho_n}{\rho_m} \le \frac{1-\beta}{\beta^2}$ & $\frac{1-\beta}{\beta^2} < \frac{\rho_n}{\rho_m} \le k_{2}$ & $\frac{\rho_n}{\rho_m} > k_{2}$ \\
        \hline
        \multirow{2}{*}{Exact results} & $P'_{T2,1}$ & $V_8$ & $V_{9}$ & 0 \\  \cline{2-5}
         & $P_{T2',2}$ & 0 & $ V_{10}$ & $V_{11}$ \\  \cline{2-5}
         & $P_{2,2}$ \cite{Sun2025HSICnPA}& $Q_9$ & $Q_9$ & $Q_{10}$ \\  \cline{2-5}
        \hline
        \multirow{2}{*}{Approximations} & $P'_{T2,1}$ & $\frac{1}{\rho_m^m} \tilde{V}_8$ & $\frac{1}{\rho_m^m} \tilde{V}_9$& 0 \\  \cline{2-5}
         & $P'_{T2,2}$ & 0 & $\frac{1}{\rho_m^m} \tilde{V}_{10}$ & $\frac{1}{\rho_m^m} \tilde{V}_{11}$ \\  \cline{2-5}
           & $P_{2,2}$ \cite{Sun2025HSICnPA}& $\frac{1}{\rho_m^m}\tilde{Q}_9$ & $\frac{1}{\rho_m^m}\tilde{Q}_9$ & $\frac{1}{\rho_m^m} \tilde{Q}_{10}$ \\  \cline{2-5}
        \hline
        \end{tabular}
    \end{center}
\label{tab4_m>n_P2}
\end{table*}

\end{theorem}
\begin{IEEEproof}
	Please refer to Appendix D.	
\end{IEEEproof}

By applying Taylor expansions, the extreme value of $\hat{P}_n$ can be obtained in the high SNR regime. Since the extreme values of $P_1$ and $P_{2,2}$ have been described in \cite{Sun2025HSICnPA}, here only presents the extreme value of $P_T$ when $m>n$, as highlighted in the following corollary.

\begin{corollary}
For the case $m>n$, when $\rho_n \to \infty$, $\rho_m \to \infty$, and $\frac{\rho_n}{\rho_m}=\eta$ is a constant, the approximations for $P_T$ can be summarized as shown in Table III and Table IV, respectively. The variables used in Table III and Table IV are listed as follows:
\begin{align}
    \tilde{V}_{1}=H_{7}(\varepsilon_m,\overline{z}_3),\quad \tilde{V}_{2}=H_{7}(\varepsilon_m,\varpi_2),
\end{align}
where 
\begin{align}
\begin{split}    
    H_7(a,b)=&\hat{c}_{mn}\!\!\!\sum_{p=0}^{m-n-1}\!\!\!\binom{m-n-1}{p}\frac{(-1)^p}{n+p}\\
    &\times \left[\Gamma_{10}(a,b)-\Gamma_{11}(a,b)\right],    
\end{split}
\end{align}
\begin{align} 
    \Gamma_{10}(a,b)\!=\!\!&\sum\limits_{{i_{_1}} + {i_2} + {i_3} = n + p}\binom{n+p}{i_{1},i_{2},i_{3}}\left(\frac{1}{\varepsilon_m}\right)^{i_1}\!\left(\frac{1}{\varepsilon_m}\!-\!1\right)^{i_2} \nonumber \\ 
\times& (-1)^{i_3}\frac{\left(b^{m-n-p+2i_1+i_2}-a^{m-n-p+2i_1+i_2}\right)}{\left(\beta\eta)^{n+p}(m-n-p+2i_1+i_2\right)},
\end{align}
\begin{align} 
\Gamma_{11}(a,b)=&\sum\limits_{q=0}^{n+p}\binom{n+p}{q}\frac{(-1)^{q}(b^{m-q}-a^{m-q})}{\left(\beta\eta\right)^{n+p}(m-q)\varepsilon_{m}^{n+p-q}}.
\end{align}
\begin{align}
    \tilde{V}_{3}=H_{8}(\overline{z}_3,\varpi_1),\quad \tilde{V}_{4}=H_{8}(\overline{z}_3,\varpi_2),
\end{align}
where 
\begin{align}
    H_8(a,b)=&{{\hat c}_{mn}}\!\!\!\sum\limits_{p = 0}^{m - n - 1}\!\!\! {\left( {\begin{array}{*{20}{c}}
			{m - n - 1}\\
			p
	\end{array}} \right)\frac{{{{( - 1)}^p}}}{{n + p}}} \\
    &\times\left[ {\frac{{b^m - a^m}}{m} - \Gamma_{11}(a,b) } \right].
\end{align}
\begin{align}
\tilde{V}_{5}=
    \begin{dcases}
    H_{9}(\varpi_{2},\overline{z}_{1}),&z_1<z_3\\
    H_{9}(\varpi_{2},\overline{z}_{3}),&z_1>z_3
    \end{dcases}
\end{align}
where
\begin{align}
\begin{split}    
    H_9(a,b)=&\hat{c}_{mn}\sum_{p=0}^{m-n-1}\binom{m-n-1}{p}\frac{(-1)^{p}}{n+p}\\
    &\times\left[\Gamma_{10}(a,b)-\Gamma_{12}(a,b)\right]    
\end{split}
\end{align}
\begin{align}
\!\!\Gamma_{12}(a,b)=&\frac{1}{\eta^{n+p}}\sum\limits_{q + j = 0}^{\infty} \sum\limits_{q = 0}^{\min (q + j, n + p)}\!\!\!\binom{n+p}{q}\binom{n+p+j-1}{j} \nonumber\\
\times&(-1)^{n+p-q}\frac{\beta^{j}}{\varepsilon_{m}^{q+j}}(b^{m-n-p+q+j}\!-\!a^{m-n-p+q+j})
\end{align}
\begin{align}
    \tilde{V}_{6}=H_{10}(\varpi_{2},\overline{z}_{2}),\quad
    \tilde{V}_{7}=
    \begin{dcases}
    0,&z_3>z_2\\
    H_{10}(\overline{z}_{3},\overline{z}_{2}), &z_3<z_2   
    \end{dcases}
\end{align}
where 
\begin{align}
\begin{split}
    H_{10}(a,b)=&\hat{c}_{mn}\sum_{p=0}^{m-n-1}\binom{m-n-1}{p}\frac{(-1)^{p}}{n+p}\\
    &\times\left[\frac{b^{m}-a^{m}}{m}-\Gamma_{12}(a,b)\right],    
\end{split}
\end{align}
\begin{align}
    \tilde{V}_{8}=H_{11}(\varepsilon_{m},\overline{z}_{3}),\quad
    \tilde{V}_{9}=
    \begin{dcases}
    H_{11}(\varepsilon_{m},\overline{z}_{3}),&z_3<\omega_4\\
    H_{11}(\varepsilon_{m},\varpi_{4}),&z_3>\omega_4
    \end{dcases}
\end{align}
where 
\begin{align}
\begin{split}    
    H_{11}(a,b)=&\hat{c}_{mn}\sum_{p=0}^{m-n-1}\binom{m-n-1}{p}\frac{(-1)^{p}}{n+p}\\
    &\times\left[\frac{b^{m}-a^{m}}{m}-\Gamma_{10}\left(a,b\right)\right].
\end{split}
\end{align}
\begin{align}
    \tilde{V}_{10}=
    \begin{dcases}
    0,&\omega_4>z_1\\
    H_{12}(\varpi_4,\overline{z}_1),&\omega_4<z_1 
    \end{dcases}, \quad
    \tilde{V}_{11}=H_{12}(\varepsilon_m,\overline{z}_1)
\end{align}
where
\begin{align}
\begin{split}
    H_{12}(a,b)=&\hat{c}_{mn}\!\!\!\sum_{p=0}^{m-n-1}\!\!\!\binom{m-n-1}{p}\frac{(-1)^{p}}{n+p}\\
    &\times\left[\Gamma_{13}(a,b)-\Gamma_{10}(a,b)\right], 
\end{split}
\end{align}
\begin{align}
\begin{split}
    \Gamma_{13}(a,b)=&\sum\limits_{q=0}^{n+p}\binom{n+p}{q}\\
    \times&\frac{(1-\beta)^{n+p-q}(1-2\beta)^{q}(b^{m-q}-a^{m-q})}{(\beta^{2}\eta)^{n+p}(m-q)}.
\end{split}
\end{align}

\begin{IEEEproof}
	Please refer to Appendix E.	
\end{IEEEproof}
\end{corollary}
{\color{black} 
\begin{Remark}
From the results in Table III and Table IV for the case $m>n$, it can be observed that the probability ${P}_T$ in the proposed H-NOMA scheme approaches 0 under all conditions, when $\rho_n \to \infty$, $\rho_m \to \infty$, and $\frac{\rho_n}{\rho_m}=\eta$ is a constant. In contrast, in the HSIC-NPA aided H-NOMA scheme, ${P}_{2,1}$ which is shown in Table III of \cite{Sun2025HSICnPA} does not approach 0 when both $\epsilon_m>\frac{\beta}{1-\beta}$ and $\frac{\rho_n}{\rho_m}> \frac{1}{\beta\epsilon_m}$ hold. 
\end{Remark}

\begin{Remark}
Since ${P}_{2,1}$ in Table III of \cite{Sun2025HSICnPA} is the inherent cause of the error floor in $\tilde{P}_n$ for HSIC-NPA aided H-NOMA scheme under certain constraints, the proposed method effectively resolves this issue. Consequently, the probability $\hat{P}_n$ approaches 0 in high SNR regime under all conditions for the case $m>n$, demonstrating the performance superiority of the proposed scheme.
\end{Remark}
}
\begin{Remark}
When $m>n$, $\rho_n \to \infty$, $\rho_m \to \infty$, and $\frac{\rho_n}{\rho_m}$ is a constant,
it is easy to observe that $\hat{P}_n$ is proportional to $\frac{1}{\rho_m^n}$ or $\frac{1}{\rho_n^n}$, which means that
$\hat{P}_n$ decays with exponential rate $n$ at high SNRs. Thus, the impact of the value of $n$ on $\hat{P}_n$ is still more dominant compared to that of the value of $m$, which is consistent with conclusions for $m<n$. 
\end{Remark}

\section{Numerical Results}
In this section, simulation results are provided to verify the accuracy of analysis and demonstrate the performance of the proposed HSIC-PA aided H-NOMA scheme. Compared with the existing HSIC-NPA aided H-NOMA scheme, the proposed scheme successfully overcomes the performance floor limitation. Moreover, it will be shown that the relative order of users $m$ and $n$ (i.e., $m > n$ or $m < n$) significantly affects the performance of the proposed scheme. Recall that the noise power is normalized to $1$, SNR in this section is same as $\rho_n$, i.e., SNR = $\rho_n$.

\begin{figure}[htbp!]
	\centering
	\vspace{-1em}
	\setlength{\abovecaptionskip}{0em}  	\setlength{\belowcaptionskip}{0em}   %
\subfloat
{\includegraphics[width=0.8\linewidth]{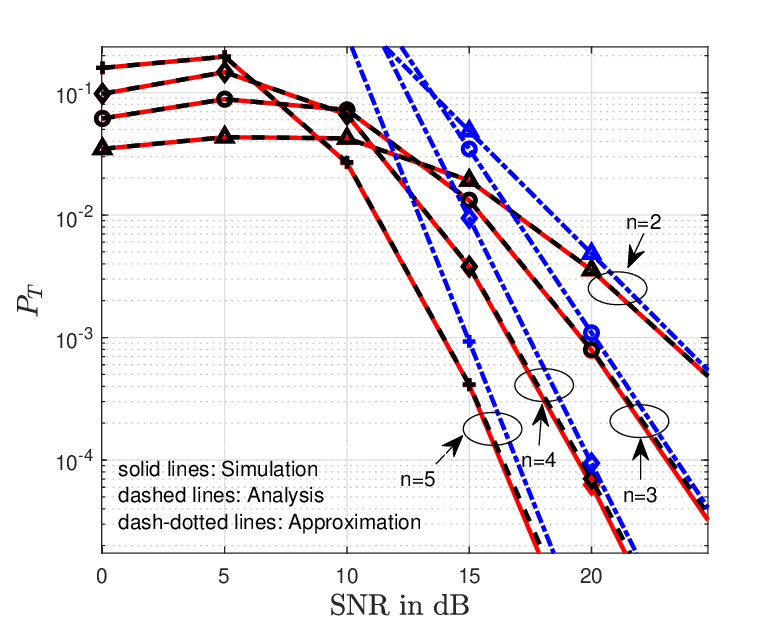}}
\caption{$P_{T}$ versus SNR for case $m < n$ with different $n$. $R_m = 0.2$ bits per channel use (BPCU), $M = 5$, $\eta={\rho_n}/{\rho_m}=1$, $m=1$ and $\beta = \frac{1}{4}$}
\label{fig1}
\end{figure}

\begin{figure}[htbp!]
	\centering
	\vspace{-1em}
	\setlength{\abovecaptionskip}{0em}  	
        \setlength{\belowcaptionskip}{0em}   %
\subfloat
{\includegraphics[width=0.8\linewidth]{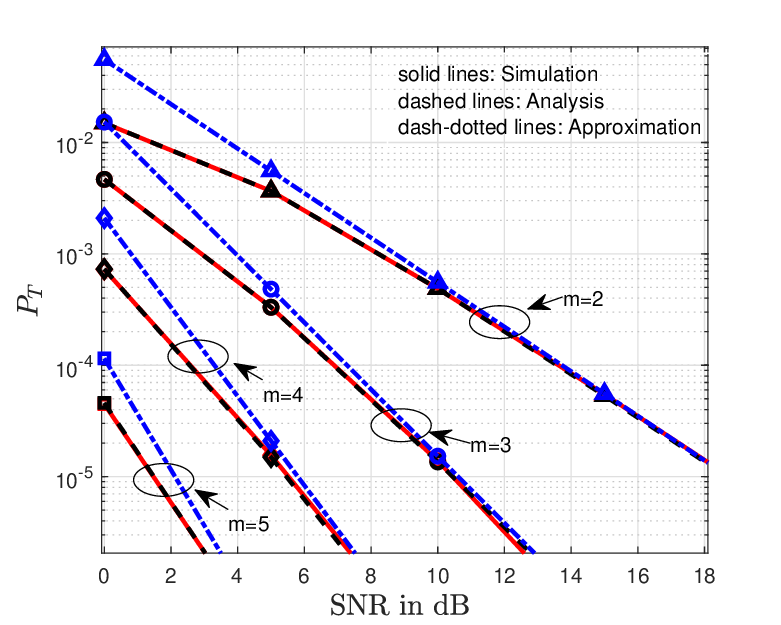} 
}%
\caption{$P_{T}$ versus SNR for case $m > n$ with different $m$. {$R_m = 0.35$ BPCU}, $M = 5$, $\eta={\rho_n}/{\rho_m}=1$, $n=1$ and $\beta = \frac{1}{4}$}
\label{fig2}
\end{figure}
Fig. \ref{fig1} and Fig. \ref{fig2} show $P_T$ versus SNR for $m<n$ and $m>n$, respectively. Note that the curves for analytical results are based on Theorems $1$ and $2$, respectively, while those for approximations are based on Corollaries $1$ and $2$. As shown in figures, the analytical results show perfect agreement with the simulation results and remain consistent with the approximate results in the high-SNR regime, verifying the accuracy of the established analysis. It can be observed that when $m<n$ with $m$ fixed, $P_T$ is significantly affected by $n$. Conversely, when $m>n$ with $n$ fixed, $P_T$ is primarily influenced by $m$. Taking the case of $m=1$ in Fig. \ref{fig1} as an example, when $n=5$, $P_T$ is the lowest among all choices of $n$ at high SNRs. Accordingly, the curve corresponding to $n=5$ has the largest slope.
\begin{figure}[ht!] 
	\centering
	\vspace{-1em}
	\setlength{\abovecaptionskip}{0em}   	
        \setlength{\belowcaptionskip}{0em}   %
\subfloat
[$\eta={\rho_n}/{\rho_m}=7$, $n=5$, $\beta=\frac{1}{3}$ ]
{\includegraphics[width=0.8\linewidth]{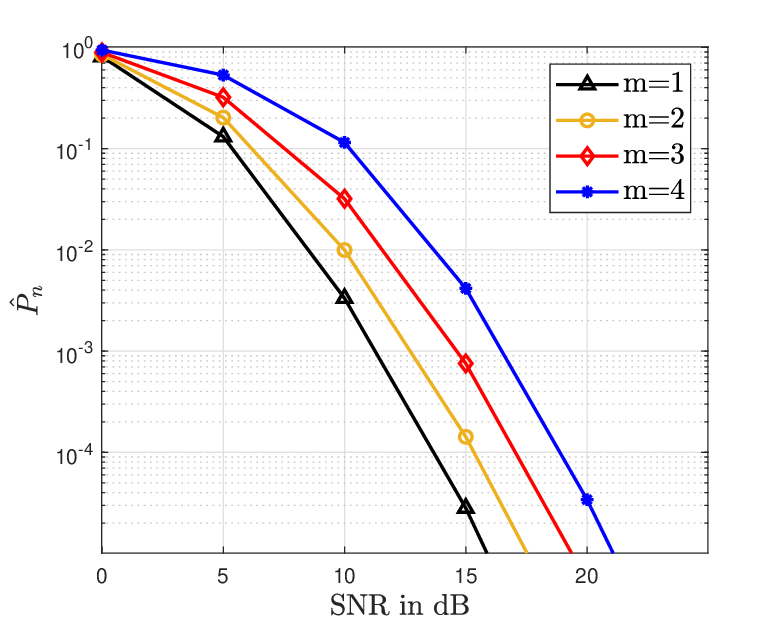}}
\label{fig3_a}
\subfloat[$\eta={\rho_n}/{\rho_m}=7$, $m=1$, $\beta=\frac{1}{3}$]
{\includegraphics[width=0.8\linewidth]{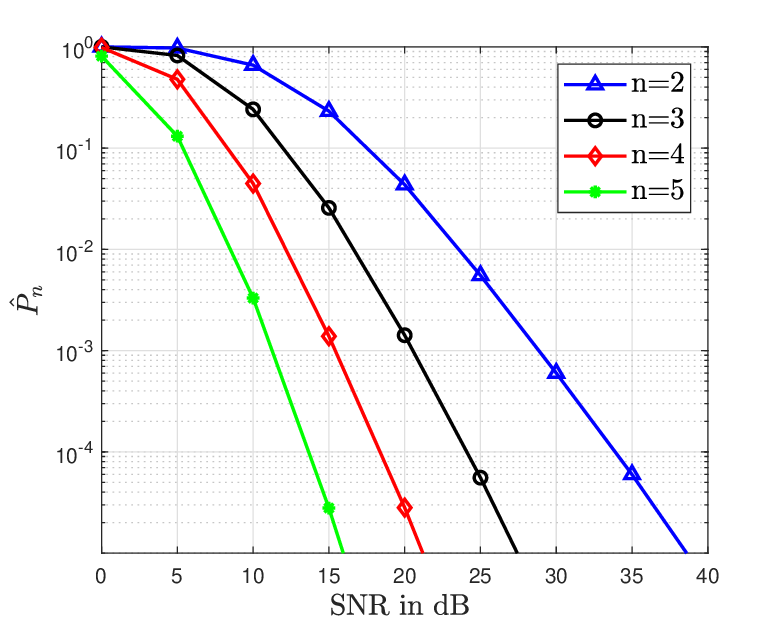}}
\label{fig3_b}
\caption{$\hat{P}_n$ with varying $m$ and varying $n$ for $m<n$. $M=5$. $R_m=1$ BPCU}
\label{fig3}
\end{figure}

\begin{figure}[ht!]
    \centering
    \vspace{-1em}
    \setlength{\abovecaptionskip}{0em}   
    \setlength{\belowcaptionskip}{0em}
\subfloat[$\eta={\rho_n}/{\rho_m}=5$, $n=1$, $\beta=\frac{1}{4}$]
{\includegraphics[width=0.8\linewidth]{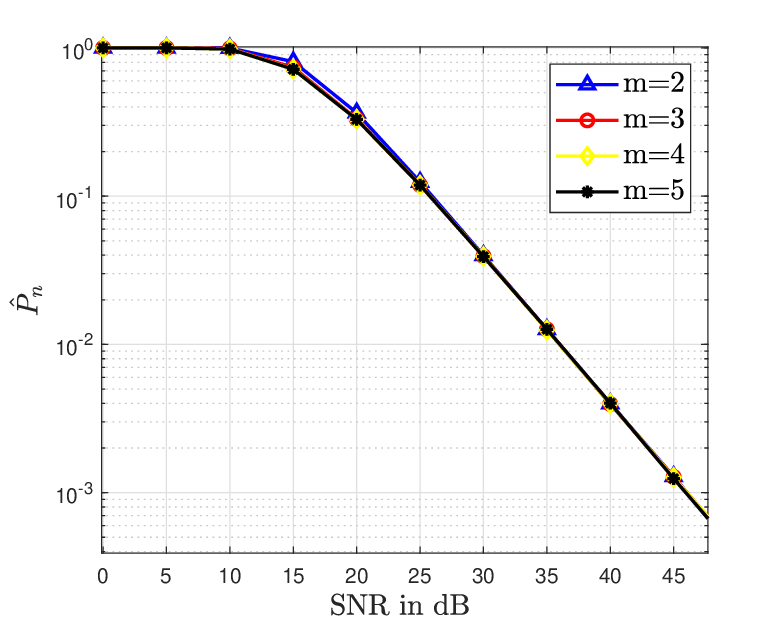}}
\label{fig4_a}
\subfloat[$\eta={\rho_n}/{\rho_m}=5$, $m=5$, $\beta=\frac{1}{4}$]
{\includegraphics[width=0.8\linewidth]{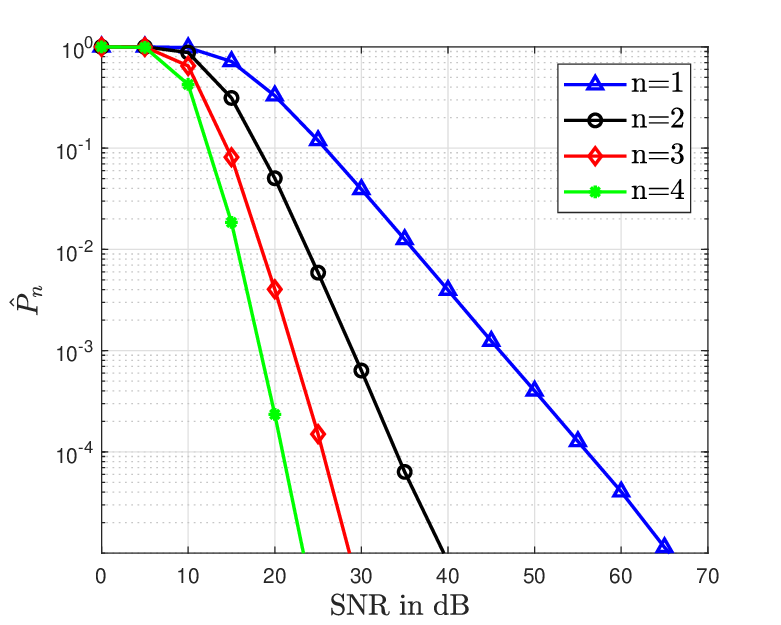}}
\label{fig4_b}
\caption{$\hat{P}_n$ with varying $m$ and varying $n$ for $m>n$. $M=5$. $R_m=1$ BPCU}
\label{fig4}
\end{figure}

\begin{figure}[ht!]
	\centering
	\vspace{-1em}
	\setlength{\abovecaptionskip}{0em}   	
        \setlength{\belowcaptionskip}{0em}   %
\subfloat[$m<n$, $m=2$, $n=5$, $\beta=\frac{1}{4}$, $R_m=1$ ]
{\includegraphics[width=0.8\linewidth]{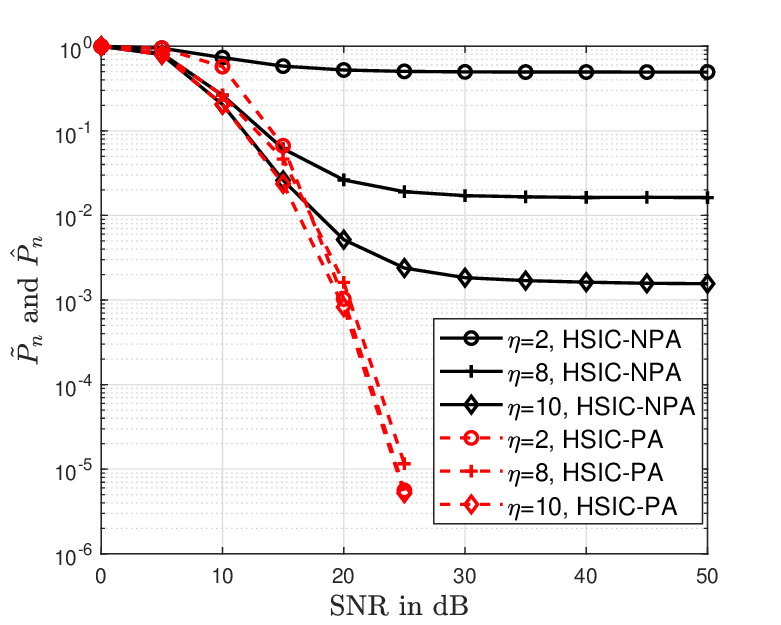}}
\label{fig5_a}
\subfloat[$m>n$, $m=3$, $n=1$, $\beta=\frac{1}{3}$, $R_m=1.5$]
{\includegraphics[width=0.8\linewidth]{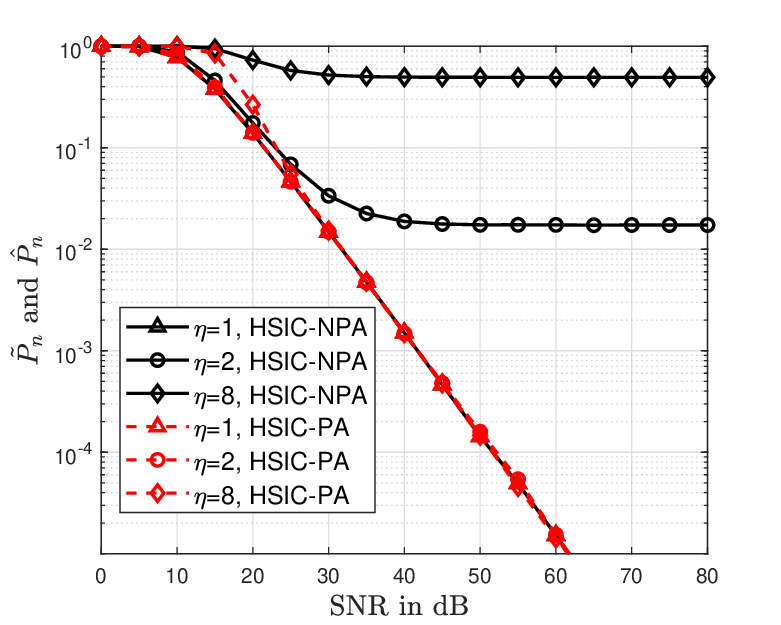}}
\label{fig5_b}
\caption{Comparisons of HSI-NPA and HSIC-PA aided hybrid NOMA schemes in terms of $\tilde{P}_n$ and $\hat{P}_n$. $M=5$, $\epsilon_m>\frac{\beta}{1-\beta}$ }
\label{fig5}
\end{figure}

Fig. \ref{fig3} shows the impact of the values of $m$ and $n$ on $\hat{P}_n$.
From Fig. \ref{fig3} (a), it can be observed that when $m<n$ with $n$ fixed, $\hat{P}_n$ decreases with $m$. However, at high SNRs, the decay rates of $\hat{P}_n$ (i.e., the slopes of the curves) for different values of $m$ are almost identical. In contrast, as seen in Fig. \ref{fig3} (b), $\hat{P}_n$ is significantly affected by $n$.
This observation is consistent with Remark $3$ that the decaying rate of $\hat{P}_n$ at high SNR is $n$, instead of $m$.

Similarly, Fig. \ref{fig4} shows the impact of $m$ and $n$ on $\hat{P}_n$ when $m>n$. From Fig. \ref{fig4} (a), it can be seen that when $n$ is fixed, the value of $m$ has little impact on $\hat{P}_n$. In contrast, as seen in Fig. \ref{fig4} (b), at high SNRs and when $m>n$, the value of $n$ still significantly affects the decay rate of $\hat{P}_n$. This observation is consistent with the conclusion in Remark $6$.

Fig. \ref{fig5} (a) and Fig. \ref{fig5}(b) compare the achieved  $\hat{P}_n$ of the proposed HSIC-PA aided H-NOMA scheme against  $\tilde{P}_n$ of the existing HSIC-NPA aided H-NOMA scheme for varying $\eta$. As illustrated in Fig. \ref{fig5} (a), there are severe floors at high SNRs for $\tilde{P}_n$ for all choices of $\eta$, while $\hat{P}_n$ of the HSIC-PA scheme can approach $0$. This is because, as stated in Remark 1, under the condition $m<n$, $\epsilon_m>\frac{\beta}{1-\beta}$, $\tilde{P}_n$ can only approach 0 when $\eta>\frac{1-\beta}{\beta^2}$. In contrast, $\hat{P}_n$ approaches 0 for all $\eta$. As depicted in Fig. \ref{fig5}(b), except for $\eta=1$, $\tilde{P}_n$ exhibits significant performance floors at high SNRs, while $\hat{P}_n$ avoids such floors for all values of $\eta$. This is because, as stated in Remark 3, when $m>n, \epsilon_m>\frac{\beta}{1-\beta}$, $\tilde{P}_n$ can only approach zero when $\eta<\frac{1}{{\beta}{\epsilon_m}}$, whereas $\hat{P}_n$ is not subject to this restriction.

\section{Conclusion}
This paper has proposed an innovative HSIC-PA aided H-NOMA scheme by introducing a power adaptation factor. Theoretical analysis established closed-form expressions for the probability $\hat{P}_{n}$ that the achievable rate of the proposed scheme cannot outperform its OMA counterpart. Both asymptotic and numerical results demonstrate the superior performance of the proposed scheme over both OMA and HSIC-NPA aided H-NOMA benchmarks, particularly in achieving $\hat{P}_{n} \to 0$ in the high SNR regime under all conditions. guaranteed performance superiority. Furthermore, user pairing studies show system performance depends more critically on the opportunistic user's order $n$ than its legacy partner $m$.

Moreover, the slow time-varying channel and perfect CSI have been assumed throughout this paper. Future researches will explore extensions to time varying channel H-NOMA scenarios and imperfect CSI conditions.
\vspace{-0.5em}
\appendices
\section{Proof for Lemma 1}
Note that $\hat{r}^{II}_n$ in \eqref{equ_rnII} can be decomposed into two forms:
\begin{align}
    \hat{r}^{II}_n\!=\! 
    \begin{dcases}\!
        1\!+\! \dfrac{\beta \rho_n \left| h_n \right|^2  }{\rho_m \left| h_m \right|^2  \!+\!1}, \!\!\!\! &\left|h_n\right|^2\!>\!\frac{(\left|h_{m}\right|^{2}\alpha_{m}^{-1}\!-\!1)(1\!+\!\rho_{m}\left|h_{m}\right|^{2})}{\beta\rho_{n}}\\
        1\!+ \!\tau_m, 
        & \left|h_n\right|^2\!<\!\frac{(\left|h_{m}\right|^{2}\alpha_{m}^{-1}\!-\!1)(1\!+\!\rho_{m}\left|h_{m}\right|^{2})}{\beta\rho_{n}}
    \end{dcases}
\end{align}
Meanwhile, $\beta\rho_n\left|h_n\right|^2>\tau_m$ can be simplified as $\left|h_n\right|^2>\frac{\left|h_{m}\right|^{2}\alpha_{m}^{-1}-1}{\beta\rho_{n}}$ under the condition $\tau_m>0$.
Thus, $P_{II,1}$ can be written as
\begin{align}
\begin{split}    
P_{{II},1} = & P\Bigg( (1+\tau_m)(1+\beta\rho_n |h_n|^2) \leq (1+\rho_n |h_n|^2),  \\
& \quad |h_n|^2 < \frac{(|h_m|^2\alpha_m^{-1}-1)(1+\rho_m|h_m|^2)}{\beta\rho_n},  \\
& \quad |h_n|^2 > \frac{|h_m|^2\alpha_m^{-1}-1}{\beta\rho_n}, \quad \tau_m > 0 \Bigg)  \\
& + P\Bigg( \left(1+\dfrac{\beta\rho_n|h_n|^2}{\rho_m|h_m|^2+1}\right)(1+\beta\rho_n|h_n|^2) \\
& \quad \quad \quad \leq (1+\rho_n|h_n|^2),  \\
& \quad |h_n|^2 > \frac{(|h_m|^2\alpha_m^{-1}-1)(1+\rho_m|h_m|^2)}{\beta\rho_n},  \\
& \quad |h_n|^2 > \frac{|h_m|^2\alpha_m^{-1}-1}{\beta\rho_n}, \quad \tau_m > 0 \Bigg)
\end{split}
\end{align}

After further simplification, the expression identical to \eqref{equ_PT} can be obtained.

\vspace{-0.5em}
\section{Proof for Theorem 1 ($m<n$)}
For the case $m<n$ (i.e., $\left|h_{m}\right|^{2}<\left|h_{n}\right|^{2}$), depending on the different magnitude relationships between the variables, the probability $P_{T1}$ and $P_{T2}$ can be expressed as the sum of probabilities, as in \eqref{equ_PT12_mln}, respectively.
\begin{figure*} 
    \begin{equation}
     \begin{split} 
     \label{equ_PT12_mln}
        P_{T1} 
    =& \underbrace{P(\Phi(\left|h_{m}\right|^{2})\!<\!\left|h_{n}\right|^{2}\! < \! \Omega(\left|h_{m}\right|^{2}), \Phi(\left|h_{m}\right|^{2}) \!>\! \left|h_{m}\right|^{2}, \Phi(\left|h_{m}\right|^{2}) \!>\! \Theta(\left|h_{m}\right|^{2}), \left|h_{m}\right|^{2} \!>\! \alpha_{m})}_{P_{T1,1}}\\
    &+
    \underbrace{P(\Theta(\left|h_{m}\right|^{2})\!<\!\left|h_n\right|^2 \!<\! \Omega(\left|h_{m}\right|^{2}), \Theta(\left|h_{m}\right|^{2}) \!>\!\left|h_m\right|^2, \Phi(\left|h_{m}\right|^{2}) \!<\!\Theta(\left|h_{m}\right|^{2}), \left|h_m\right|^2 \!>\! \alpha_m)}_{P_{T1,2}}\\
    &+
    \underbrace{P(\left|h_m\right|^2 \!<\! \left|h_n\right|^2 \!<\! \Omega(\left|h_{m}\right|^{2}), \left|h_m\right|^2 \!>\! \Phi(\left|h_{m}\right|^{2}), \left|h_m\right|^2 \!>\! \Theta(\left|h_{m}\right|^{2}), \left|h_m\right|^2 \!>\! \alpha_m)}_{P_{T1,3}}, \\
    P_{T2} 
    =&\underbrace{P(\left|h_{m}\right|^{2}<\left|h_{n}\right|^{2}<\Psi(\left|h_{m}\right|^{2}), 
    \left|h_{m}\right|^{2}>\Omega(\left|h_{m}\right|^{2}),
    \left|h_{m}\right|^{2}>\alpha_{m})}_{P_{T2,1}} \\
    &+ 
    \underbrace{P(\Omega(\left|h_{m}\right|^{2})<\left|h_{n}\right|^{2}<\Psi(\left|h_{m}\right|^{2}), 
    \left|h_{m}\right|^{2}<\Omega(\left|h_{m}\right|^{2}),
    \left|h_{m}\right|^{2}>\alpha_{m})}_{P_{T2,2}}.    
    \end{split}
\end{equation}
\vspace{-1.5em}
\end{figure*}  

As previously mentioned, $P_{T}=P_{T1,1}+P_{T1,2}+P_{T1,3}+P_{T2,1}+P_{T2,2}$. Since the proofs of the other parts are similar to that of $P_{T1,2}$, only the proof for $P_{T1,2}$ is provided here.
Under the condition $\left|h_{m}\right|^{2}<\left|h_{n}\right|^{2}$ and $\frac{\rho_n}{\rho_m}\le k_1$, $P_{T1,2}$ can be rewritten as follows:
\begin{align}
\begin{split}
P_{T1,2}=&P(\Theta(\left|h_{m}\right|^{2}<\left|h_n\right|^2<\Omega(\left|h_{m}\right|^{2}), \\
& \quad \omega_2<\left|h_{m}\right|^{2}<z_1).    
\end{split}
\end{align}
Note that the users are ordered according to their channel gains, hence the joint probability density function (pdf) of $|h_m|^2$ and $|h_n|^2$ $(m<n)$ is given by:
 \begin{align}
 \begin{split}
&f_{|h_{m}|^2,|h_n|^2}(x,y)\\
&=c_{mn}\!\!\!\sum_{p=0}^{n-m-1}\!\!\!\!c_p\!\!\sum_{l=0}^{m-1}\!c_l e^{-(l+p+1)x}e^{-(M-m-p)y}   
 \end{split}
 \end{align}
Hence, when $\frac{\rho_n}{\rho_m}\le k_1$, $P_{T1,2}$ can be expressed as
\begin{align}
\begin{split}
P_{T1,2}=&\int_{\omega_2}^{z_1}\int_{\Theta(x)}^{\Omega(x)}
 c_{mn}\!\!\!\!\sum_{p=0}^{n-m-1}\!\!\!\!c_p\!\!\sum_{l=0}^{m-1}\!c_l \\
 &\times e^{-(l+p+1)x} 
 e^{-(M-m-p)y}dydx\\
 =&{c_{mn}}\!\!\!\!\sum\limits_{p = 0}^{n - m - 1} \!\!\!{{c_p}\sum\limits_{l = 0}^{m - 1} \!{{c_l}} } \frac{1}{{M - m - p}}\frac{{{z_1} - {w_2}}}{2}\\
 & \times \frac{\pi }{{{n_c}}}\sum\limits_{i = 1}^{{n_c}} \Gamma_2(\frac{z_{1}-w_{2}}{2}t_{i}+\frac{z_{1}+w_{2}}{2})\sqrt{1-t_{i}^{2}}.
\end{split}
\end{align}
Here the Gauss-Chebyshev quadrature formula is applied:
\begin{align} \label{equ_GaussCheb}
\begin{split}
    \int_{a}^{b}f(x)dx&=\sum_{i=1}^{n_c}\frac{\pi}{n_c}(\frac{a}{2}-\frac{b}{2})\\
    &\times f[(\frac{a}{2}+\frac{b}{2})+(\frac{a}{2}-\frac{b}{2})t_i]\sqrt{1-t_i^{2}}.    
\end{split}
\end{align}
And $n_c$ is parameter for Gauss-Chebyshev approximation, $t_{i}=\cos(\frac{2i-1}{2n_{c}}\pi)$.

Accordingly, $P_{T1,2}$ can be expressed as $S_2$ when $\left|h_{m}\right|^{2}<\left|h_{n}\right|^{2}$ and $\frac{\rho_n}{\rho_m}\le k_1$.

\vspace{-0.5em}
\section{Proof for Corollary 1 ($m<n$)}
Considering that the proof steps for other probability are similar, only the proof of the approximation for $P_{T1,2}$ is provided below.

To facilitate the approximation,  the pdf of $|h_m|^2$ and $|h_n|^2$ $(m<n)$, derived in \cite{Sun2025HSICnPA}, can be expressed as:
\begin{align}
\!f_{|h_{m}|^{2},|h_{n}|^{2}}(x,y)=&c_{mn}\left(F(x)\right)^{m\!-\!1}\left(F(y)\!-\!F(x)\right)^{n\!-\!m\!-\!1} \nonumber \\
&\times \left(1-F(y)\right)^{M-n}f(x)f(y),
\end{align}
where $x < y$, $f(x)=e^{-x}$ and $F(x)=1-e^{-x}$. 
Using Taylor series expansion, we obtain the approximations $f(x)=e^{-x}\approx 1$ and $F(x)=1-e^{-x} \approx x$ when $x\to 0$. 
Furthermore, by applying the binomial expansion theorem, $f_{|h_{m}|^{2},|h_{n}|^{2}}(x,y)$ can be further approximated as
\begin{align}
\begin{split}    
f_{|h_{m}|^{2},|h_{n}|^{2}}(x,y) \approx & c_{mn}\sum_{p=0}^{n-m-1}\binom{n-m-1}{p}\\
&\times(-1)^{p}y^{n-m-1-p}x^{m-1+p}.
\end{split}
\end{align}

Employing the above pdf, $P_{T1,2}$ can be rewritten as 
\begin{align}\label{P_{T1,2}}
P_{T1,2}=&\int_{\omega_2}^{z_1}\int_{\Theta(x)}^{\Omega(x)}
 c_{mn}\!\!\!\!\sum_{p=0}^{n-m-1}\!\!\!\binom{n-m-1}{p} \\ 
 \times & (-1)^{p}y^{n-m-1-p}x^{m-1+p}dxdy \nonumber\\
 =&\frac{1}{\rho_m^n}c_{mn}\!\!\!\!\sum_{p=0}^{n-m-1}\!\!\!\binom{n-m-1}{p}\frac{\left(-1\right)^{p}}{n-m-p} \nonumber\\
 \Bigg[&\sum\limits_{i_{1}+i_{2}+i_{3}=n-m-p}\!\!\binom{n\!-\!m\!-\!p}{i_{1},i_{2},i_{3}} (\frac{1}{\varepsilon_{m}})^{i_1}(\frac{1}{\varepsilon_{m}}\!-\!1)^{i_{2}} \nonumber\\
 \times & (-1)^{i_{3}}
\frac{\left(\bar{z}_1^{m+p+2i_{1}+i_{2}}-\varpi_2^{m+p+2i_{1}+i_{2}}\right)}{\left(\beta\eta)^{n-m-p}(m+p+2i_{1}+i_{2}\right)}\nonumber\\
-&\frac{1}{\eta^{n-m-p}}\sum\limits_{q+j=0}^{\infty} \sum\limits_{q=0}^{\min (q+j, n-m-p)}\binom{n-m-p}{q}\nonumber\\
\times&\binom{n-m-p+j-1}{j}(-1)^{n-m-p-q}\nonumber\\
\times&\frac{\beta^{j}}{\varepsilon_{m}^{q+j}}(\bar{z}_1^{m+p+q+j}-\varpi_2^{m+p+q+j})\Bigg] \nonumber
\end{align}
The multinomial theorem, the binomial theorem and the properties of power series are applied there, as shown below.  
\begin{align}
\begin{split}
\!\!(x_1\!+\!x_2\!+\!\cdots\!+\!x_m)^n=\!\!\!\!\sum_{k_1+k_2+\cdots+k_m=n}&\frac{n!}{k_1!k_2!\cdots k_m!}\\
\times & x_1^{k_1}x_2^{k_2}\cdots x_m^{k_m}       
\end{split}
\end{align}
\begin{align}
\begin{split}    
&\left(\!\frac{ax-1}{1-bx}\!\right)^{n-m-p}\\
=&\sum_{q+j=0}^\infty\Bigg(\sum_{q=0}^{\min(q+j,n-m-p)}\binom{n-m-p}{q}\\
&\times \binom{n-m-p+j-1}{j}(-1)^{n-m-p-q}a^qb^j\Bigg)x^{q+j}
\end{split}
\end{align}
Accordingly, $P_{T1,2}$ can be approximated by $\frac{1}{\rho_m^n}\tilde{S_2}$ in the high-SNR regime.

\vspace{-0.5em}
\section{Proof for Theorem 2 ($m>n$)}
For the case $m>n$ (i.e., $\left|h_{m}\right|^{2}>\left|h_{n}\right|^{2}$), depending on the different magnitude relationships between the variables, the probability $P_{T1}$ and $P_{T2}$ can be expressed as the sum of probabilities, as shown in \eqref{equ_PT12_mgn}, respectively.
\begin{figure*}  
\begin{equation} 
    \begin{split} 
    \label{equ_PT12_mgn}
    P_{T1}
    =& \underbrace{P(\Phi(\left|h_{m}\right|^{2})<\left|h_{n}\right|^{2}<\Omega(\left|h_{m}\right|^{2}), \Phi(\left|h_{m}\right|^{2})>\Theta(\left|h_{m}\right|^{2}), 
    \left|h_{m}\right|^{2}>\Omega(\left|h_{m}\right|^{2}), 
    \left|h_{m}\right|^{2}>\alpha_{m})}_{P'_{T1,1}} \\
    &+
    \underbrace{P(\Phi(\left|h_{m}\right|^{2})<\left|h_{n}\right|^{2}<\left|h_{m}\right|^{2}, \Phi(\left|h_{m}\right|^{2})>\Theta(\left|h_{m}\right|^{2}), 
    \left|h_{m}\right|^{2}<\Omega(\left|h_{m}\right|^{2}), 
    \left|h_{m}\right|^{2}>\alpha_{m})}_{P'_{T1,2}} \\
    &+
    \underbrace{P(\Theta(\left|h_{m}\right|^{2})<\left|h_{n}\right|^{2}<\Omega(\left|h_{m}\right|^{2}), 
    \Phi(\left|h_{m}\right|^{2})<\Theta(\left|h_{m}\right|^{2}), 
    \left|h_{m}\right|^{2}>\Omega(\left|h_{m}\right|^{2}),
    \left|h_{m}\right|^{2}>\alpha_{m})}_{P'_{T1,3}} \\
    &+
    \underbrace{P(\Theta(\left|h_{m}\right|^{2})<\left|h_{n}\right|^{2}<\left|h_{m}\right|^{2}, 
    \Phi(\left|h_{m}\right|^{2})<\Theta(\left|h_{m}\right|^{2}), 
    \left|h_{m}\right|^{2}<\Omega(\left|h_{m}\right|^{2}),
    \left|h_{m}\right|^{2}>\alpha_{m})}_{P'_{T1,4}}\\
    P_{T2} 
    =& \underbrace{P(\Omega(\left|h_{m}\right|^{2})<\left|h_n\right|^2<\left|h_m\right|^2, 
    \left|h_m\right|^2<\Psi(\left|h_{m}\right|^{2}), 
    \left|h_m\right|^2>\alpha_m)}_{P'_{T2,1}}\\
    &+
    \underbrace{P(\Omega(\left|h_{m}\right|^{2})<\left|h_n\right|^2<\Psi(\left|h_{m}\right|^{2}),
    \left|h_m\right|^2>\Psi(\left|h_{m}\right|^{2}),
    \left|h_m\right|^2>\alpha_m)}_{P'_{T2,2}}.
    \end{split}
\end{equation}
\vspace{-1.5em}
\end{figure*}   
In this section, we present a detailed proof for $P'_{T1,1}$ specifically, while the other components can be proved similarly.
Note that $\Phi(\left|h_{m}\right|^{2})<\Omega(\left|h_{m}\right|^{2})$ always holds. When $\left|h_{m}\right|^{2}<\omega_2$, $\Phi(\left|h_{m}\right|^{2})>\Theta(\left|h_{m}\right|^{2})$ holds, and when $\left|h_{m}\right|^{2}<z_3$, $\left|h_{m}\right|^{2}>\Omega(\left|h_{m}\right|^{2})$ holds. Meanwhile, note that if $\frac{\rho_n}{\rho_m}<k_1$, $\alpha_m<z_3<\omega_2$ holds. Thus when $\frac{\rho_n}{\rho_m}\le k_1$, $P'_{T1,1}$ can be rewritten as follows:
\begin{align}
\begin{split}
P'_{T1,1}=&P(\Phi(\left|h_{m}\right|^{2})<\left|h_{n}\right|^{2}<\Omega(\left|h_{m}\right|^{2}),\\
&\quad \alpha_m<\left|h_{m}\right|^{2}<z_3)
\end{split}
\end{align}
Since the users are ordered according to their channel gains,
the joint pdf of $|h_n|^2$ and $|h_m|^2$ $(m>n)$ is given by:
\begin{align}
\begin{split}
 &f_{|h_{n}|^2,|h_m|^2}(x,y)\\
 &=\hat{c}_{mn}\!\!\!\sum_{p=0}^{m-n-1}\!\!\!\!\hat{c}_p\!\!\sum_{l=0}^{n-1}\!\hat{c}_l e^{-(l+p+1)x}e^{-(M-n-p)y}.   
\end{split}
\end{align}
Thus, when $\frac{\rho_n}{\rho_m}\le k_1$, $P'_{T1,1}$ can be expressed as :
\begin{align}
P'_{T1,1}=&\int_{\alpha_m}^{z_3}\int_{\Phi(y)}^{\Omega(y)}
 f_{|h_{n}|^2,|h_m|^2}(x,y)dx dy\nonumber\\
=&\hat{c}_{mn}\sum_{p=0}^{m-n-1}\hat{c}_{p}\sum_{l=0}^{n-1}\hat{c}_{l}\frac{e^{\frac{l+p+1}{\beta\rho_{n}}}}{l+p+1} \Bigg[\frac{e^{-r_3\alpha_{m}}-e^{-r_3z_{3}}}{r_3}\nonumber\\
&-\Gamma_1(\alpha_m,z_3,D,E)
\end{align}
Hence, for the case $\frac{\rho_n}{\rho_m}\!\le\! k_1$, $P'_{T1,1}$ can be expressed as $V_1$.

\vspace{-0.5em}
\section{Proof for Corollary 2 ($m>n$)}
Noting that the approximation procedures are similar, the next section will provide only the proof for $P'_{T1,1}$ when $\frac{\rho_n}{\rho_m}\le k_1$, due to limited space.

Recall that when $\frac{\rho_n}{\rho_m}\le k_1$, $P'_{T1,1}$ can be expressed as follows:
\begin{align}
\begin{split}
P'_{T1,1}=&P(\Phi(\left|h_{m}\right|^{2})<\left|h_{n}\right|^{2}<\Omega(\left|h_{m}\right|^{2}),\\
& \quad \alpha_m<\left|h_{m}\right|^{2}<z_3)    
\end{split}
\end{align}
In order to facilitate approximation, the pdf of $|h_n|^2$ and $|h_m|^2$ ($m>n$) can be rewritten as follows:
\begin{align}
f_{|h_{n}|^2,|h_m|^2}(x,y)=&\hat{c}_{mn}\left( F(x) \right)^{n\!-\!1}\left(  F(y) \!-\!F(x) \right)^{m\!-\!n\!-\!1} \nonumber\\
&\times\left( 1\!-\!F(y) \right)^{M\!-\!m}f(x)f(y)
\end{align}
By applying Taylor seizes and the binomial expansion theorem, $f_{|h_{n}|^{2},|h_{m}|^{2}}(x,y)$ can be further approximated as 
\begin{align}
\begin{split}    
f_{|h_{n}|^{2},|h_{m}|^{2}}(x,y)\approx \hat{c}_{mn}\!\!\sum_{p=0}^{m-n-1}\binom{m-n-1}{p}\\ \times (-1)^{p}y^{m-n-1-p}x^{n-1+p}. 
\end{split}
\end{align}
When $\frac{\rho_n}{\rho_m}\le k_1$, $P'_{T1,1}$ can be rewritten as follows:
\begin{align}
P'_{T1,1}=&\int_{\alpha_m}^{z_3}\int_{\Phi(y)}^{\Omega(y)}
(-1)^{p}y^{m-n-1-p}x^{n-1+p}dx dy \nonumber\\
=&\frac{1}{\rho_m^m}\hat{c}_{mn}\!\!\!\sum_{p=0}^{m-n-1}\!\!\binom{m-n-1}{p}\frac{(-1)^p}{n+p} \nonumber\\
    \!\!\Bigg[&\sum\limits_{{i_{_1}} + {i_2} + {i_3} = n + p}\!\!\binom{n+p}{i_{1},i_{2},i_{3}}\!\left(\frac{1}{\varepsilon_m}\!\right)^{i_1}\!\left(\!\frac{1}{\varepsilon_m}\!-\!1\!\right)^{i_2}\\
    \times & (-1)^{i_3} \frac{\left(\bar{z}_3^{m-n-p+2i_1+i_2}-\varepsilon_m^{m-n-p+2i_1+i_2}\right)}{\left(\beta\eta)^{n+p}(m-n-p+2i_1+i_2\right)} \nonumber\\
-&\sum\limits_{q=0}^{n+p}\binom{n+p}{q}\frac{(-1)^{q}(\bar{z}_3^{m-q}-\varepsilon_m^{m-q})}{\left(\beta\eta\right)^{n+p}(m-q)\varepsilon_{m}^{n+p-q}}\Bigg] \nonumber
\end{align}
Hence, when $\frac{\rho_n}{\rho_m}\!\le\! k_1$, $P'_{T1,1}$ can be approximated as $\frac{1}{\rho_m^m}\tilde{V}_1$.

\vspace{-0.5em}
\bibliographystyle{IEEEtran}
\bibliography{IEEEabrv, ref}

\begin{thebibliography}{10}
\providecommand{\url}[1]{#1}
\csname url@samestyle\endcsname
\providecommand{\newblock}{\relax}
\providecommand{\bibinfo}[2]{#2}
\providecommand{\BIBentrySTDinterwordspacing}{\spaceskip=0pt\relax}
\providecommand{\BIBentryALTinterwordstretchfactor}{4}
\providecommand{\BIBentryALTinterwordspacing}{\spaceskip=\fontdimen2\font plus
\BIBentryALTinterwordstretchfactor\fontdimen3\font minus \fontdimen4\font\relax}
\providecommand{\BIBforeignlanguage}[2]{{%
\expandafter\ifx\csname l@#1\endcsname\relax
\typeout{** WARNING: IEEEtran.bst: No hyphenation pattern has been}%
\typeout{** loaded for the language `#1'. Using the pattern for}%
\typeout{** the default language instead.}%
\else
\language=\csname l@#1\endcsname
\fi
#2}}
\providecommand{\BIBdecl}{\relax}
\BIBdecl

\bibitem{dai2018survey}
L.~Dai, B.~Wang, Z.~Ding, Z.~Wang, S.~Chen, and L.~Hanzo, ``{A survey of non-orthogonal multiple access for 5G},'' \emph{{IEEE} Commun. Surveys Tuts.}, vol.~20, no.~3, pp. 2294--2323, May 2018.

\bibitem{makki2020survey}
B.~Makki, K.~Chitti, A.~Behravan, and M.-S. Alouini, ``{A survey of NOMA: Current status and open research challenges},'' \emph{IEEE Open Journal of the Commun. Society}, vol.~1, pp. 179--189, 2020.

\bibitem{you2021towards}
X.~You, C.-X. Wang, J.~Huang, X.~Gao, Z.~Zhang, M.~Wang, Y.~Huang, C.~Zhang, Y.~Jiang, J.~Wang \emph{et~al.}, ``{Towards 6G wireless communication networks: Vision, enabling technologies, and new paradigm shifts},'' \emph{{Sci. China Inf. Sci. }}, vol.~64, pp. 1--74, Feb. 2021.

\bibitem{sun2024study}
L.~Sun, Z.~Zhao, S.~Wang, Z.~Ding, and M.~Peng, ``On the study of non-orthogonal multiple access (noma)-assisted integrated sensing and communication (isac),'' \emph{IEEE Transactions on Communications}, vol.~72, no.~11, pp. 7278--7293, 2024.

\bibitem{li2023achievable}
Q.~Li, M.~El-Hajjar, Y.~Sun, I.~Hemadeh, A.~Shojaeifard, Y.~Liu, and L.~Hanzo, ``{Achievable rate analysis of the STAR-RIS-aided NOMA uplink in the face of imperfect CSI and hardware impairments},'' \emph{IEEE Transactions on Communications}, vol.~71, no.~10, pp. 6100--6114, Oct. 2023.

\bibitem{zhu2020power}
J.~Zhu, Y.~Huang, J.~Wang, K.~Navaie, and Z.~Ding, ``Power efficient irs-assisted noma,'' \emph{IEEE Transactions on Communications}, vol.~69, no.~2, pp. 900--913, Feb. 2020.

\bibitem{mu2023exploiting}
X.~Mu and Y.~Liu, ``{Exploiting semantic communication for non-orthogonal multiple access},'' \emph{{IEEE} J. Select. Areas Commun.}, vol.~41, no.~8, pp. 2563--2576, Aug. 2023.

\bibitem{Ding2025Pinching}
Z.~Ding, R.~Schober, and H.~Vincent~Poor, ``Flexible-antenna systems: A pinching-antenna perspective,'' \emph{IEEE Transactions on Communications}, 2025, early access: doi={10.1109/TCOMM.2025.3555866}.

\bibitem{Yang2025Pinching}
Z.~Yang, N.~Wang, Y.~Sun, Z.~Ding, R.~Schober, G.~K. Karagiannidis, V.~W.~S. Wong, and O.~A. Dobre, ``Pinching antennas: Principles, applications and challenges,'' \emph{(submitted)}, arXiv:2501.10753.

\bibitem{Liu2025Pinching}
Y.~Liu, Z.~Wang, X.~Mu, C.~Ouyang, X.~Xu, and Z.~Ding, ``Pinching-antenna systems (pass): Architecture designs, opportunities, and outlook,'' \emph{(submitted)}, arXiv:2501.18409.

\bibitem{Ding2019MEC}
Z.~{Ding}, P.~{Fan}, and H.~V. {Poor}, ``{Impact of non-orthogonal multiple access on the offloading of mobile edge computing},'' \emph{{IEEE} Trans. Commun.}, vol.~67, no.~1, pp. 375--390, Jan. 2019.

\bibitem{liu2021latency}
L.~Liu, B.~Sun, Y.~Wu, and D.~H. Tsang, ``{Latency optimization for computation offloading with hybrid NOMA--OMA transmission},'' \emph{IEEE Internet of Things J.}, vol.~8, no.~8, pp. 6677--6691, Aug. 2021.

\bibitem{wei2022energy}
X.~Wei, H.~Al-Obiedollah, K.~Cumanan, Z.~Ding, and O.~A. Dobre, ``{Energy efficiency maximization for hybrid TDMA-NOMA system with opportunistic time assignment},'' \emph{{{IEEE} Trans. Veh. Technol.}}, vol.~71, no.~8, pp. 8561--8573, Aug. 2022.

\bibitem{Ding2025HNOMAopt}
Z.~Ding, ``A study on the optimality of downlink hybrid noma,'' \emph{IEEE Signal Processing Letters}, vol.~32, pp. 511--515, 2025.

\bibitem{Fang2025Rethinking}
X.~Xie, F.~Fang, and X.~Wang, ``Rethinking power minimization in a downlink hybrid noma network,'' \emph{IEEE Communications Letters}, vol.~29, no.~5, pp. 953--957, 2025.

\bibitem{higuchi2013non}
K.~Higuchi and Y.~Kishiyama, ``{Non-orthogonal access with random beamforming and intra-beam SIC for cellular MIMO downlink},'' in \emph{IEEE Veh. Tech. Conf.,}, Las Vegas, NV, US, Sep. 2013, pp. 1--5.

\bibitem{gao2017theoretical}
Y.~{Gao}, B.~{Xia}, K.~{Xiao}, Z.~{Chen}, X.~{Li}, and S.~{Zhang}, ``Theoretical analysis of the dynamic decode ordering sic receiver for uplink noma systems,'' \emph{{IEEE} Commun. Lett.}, vol.~21, no.~10, pp. 2246--2249, Jun. 2017.

\bibitem{Xia2018outage}
B.~{Xia}, J.~{Wang}, K.~{Xiao}, Y.~{Gao}, Y.~{Yao}, and S.~{Ma}, ``{Outage Performance Analysis for the Advanced SIC Receiver in Wireless NOMA Systems},'' \emph{{IEEE} Trans. Veh. Technol.}, vol.~67, no.~7, pp. 6711--6715, Mar. 2018.

\bibitem{zhou2018state}
F.~Zhou, Y.~Wu, Y.-C. Liang, Z.~Li, Y.~Wang, and K.-K. Wong, ``{State of the art, taxonomy, and open issues on cognitive radio networks with NOMA},'' \emph{IEEE Wireless Commun.}, vol.~25, no.~2, pp. 100--108, 2018.

\bibitem{Dhakal2019noma}
S.~{Dhakal}, P.~A. {Martin}, and P.~J. {Smith}, ``{NOMA With Guaranteed Weak User QoS: Design and Analysis},'' \emph{IEEE Access}, vol.~7, pp. 32\,884--32\,896, Mar. 2019.

\bibitem{ding2021new}
Z.~Ding, R.~Schober, and H.~V. Poor, ``{A new QoS-guarantee strategy for NOMA assisted semi-grant-free transmission},'' \emph{IEEE Trans. Commun.}, vol.~69, no.~11, pp. 7489--7503, Jul. 2021.

\bibitem{ding2020unveiling1}
------, ``Unveiling the importance of sic in noma systems—part 1: State of the art and recent findings,'' \emph{IEEE Communications Letters}, vol.~24, no.~11, pp. 2373--2377, 2020.

\bibitem{ding2020unveiling2}
------, ``Unveiling the importance of sic in noma systems—part ii: New results and future directions,'' \emph{IEEE Communications Letters}, vol.~24, no.~11, pp. 2378--2382, 2020.

\bibitem{Yang2023HSICpower}
Z.~Yang, J.~A. Hussein, P.~Xu, G.~Chen, Y.~Wu, and Z.~Ding, ``A novel hybrid successive interference cancellation for uplink wireless power transfer noma in internet of things,'' \emph{IEEE Transactions on Vehicular Technology}, vol.~72, no.~5, pp. 6090--6102, 2023.

\bibitem{Sun2025HSICnPA}
Y.~Sun, W.~Cao, N.~Wang, M.~Zhou, and Z.~Ding, ``Hybrid sic-aided hybrid noma: A new approach for improving energy efficiency,'' \emph{IEEE Transactions on Wireless Communications}, vol.~24, no.~3, pp. 2007--2021, 2025.

\bibitem{sun2021new}
Y.~Sun, Z.~Ding, and X.~Dai, ``{A new design of hybrid SIC for improving transmission robustness in uplink NOMA},'' \emph{{IEEE} Trans. Veh. Technol.}, vol.~70, no.~5, pp. 5083--5087, Mar. 2021.

\bibitem{Sun2023Globecom}
Y.~Sun, W.~Cao, M.~Zhou, and Z.~Ding, ``{New designs of robust uplink NOMA in cognitive radio inspired communications},'' in \emph{2023 IEEE Globecom Workshops}, 2023, pp. 1207--1212.

\bibitem{sun2023hybrid}
------, ``{Hybrid successive interference cancellation and power adaptation: a win-win strategy for robust uplink NOMA transmission},'' \emph{IEEE Trans. Commun.}, vol.~72, no.~2, pp. 771--785, Feb. 2024.

\end{thebibliography}
\end{document}